\documentclass[a4paper,12pt]{article}
\usepackage{amsmath, amssymb, graphicx, physics, bm, mathrsfs}
\usepackage[dvipsnames,svgnames,table,x11names]{xcolor}
\usepackage[a4paper, margin=1in]{geometry}
\usepackage{tikz}
\usepackage{subcaption}
\usetikzlibrary{arrows.meta,calc}
\usetikzlibrary{decorations.markings, arrows.meta}
\usepackage{jheppub, appendix, mathrsfs}

\title{\boldmath Spectral Topology and Universal Krylov Dynamics}

\author{Jeff Murugan$^{1}$, Hendrik J. R. Van Zyl$^{1}$ \& Masataka Watanabe$^{2}$}
\affiliation{$^{1}$The Laboratory for Quantum Gravity \& Strings,\\
Department of Mathematics \& Applied Mathematics,\\
University of Cape Town,\\ Rondebosch, Cape Town, 7701,\\ South Africa\\}
\affiliation{$^{2}$Faculty of Science,\\ The University of Tokyo,\\ 
Tokyo 113-0033, Japan}

\emailAdd{jeff.murugan@uct.ac.za}

\abstract{The leading asymptotic growth of Lanczos coefficients is controlled by spectral tails and furnishes a coarse classification of Krylov dynamics. We show that the \textit{global topology} of the spectral measure, specifically the number of connected components, the gap structure, and the behaviour at gap-closing transitions, encodes a finer hierarchy of dynamical invariants invisible to tail-based arguments. Using the Riemann-Hilbert formulation of orthogonal polynomials and Deift-Zhou steepest descent, we recover the Freud growth laws $b_n\sim n^{1/\beta}$ for single-cut measures and determine their sub-leading corrections from endpoint data. Gapped spectra produce quasiperiodic Lanczos oscillations at a frequency fixed by the filling fraction of the spectral bands alone, and hence predictable from the band edges. We verify this in the SSH chain and its next-nearest-neighbour deformation. At a gap-closing transition the oscillation amplitude is governed by the Hastings-McLeod solution of Painlevé II, decaying as $n^{-1/3}$ at criticality and interpolating between the gapped and merged phases, so that the topology change of the spectral curve is realised as a Krylov phase transition. We also demonstrate that, while in the conformal limit of SYK the operator scaling dimension is invisible in the leading rate $\alpha = \pi T$, it can be extracted from the subleading offset $b_0 = \pi T(\Delta - \frac{1}{2})$. These results establish a refined notion of universality in operator growth, classified by spectral topology rather than spectral tails alone. }

\dedicated{Dedicated to the memory of Kathy Driver.}

\begin{document}
\maketitle
\flushbottom

\section{Introduction}
\label{sec:intro}
There are few things more satisfying to a theoretical physicist than finding that some arcane corner of mathematics, often encountered in passing as an undergraduate or graduate student, is exactly what is needed to understand a completely unrelated physics problem years\footnote{Or, for at least one of us, decades!} later. By recasting Heisenberg evolution as a quantum walk on a semi-infinite tight-binding chain, Krylov complexity \cite{Parker:2018yvk, Balasubramanian:2022tpr, Nandy:2024evd, Rabinovici:2025otw, Baiguera:2025dkc} provides a geometric picture of information scrambling in which the dynamics is governed by a sequence of Lanczos coefficients, ${b_n}$ \cite{Lanczos1950AnIM}. Over the past few years, these coefficients have become recognized as fundamental dynamical observables, encoding aspects of ergodicity, chaos, thermalization and operator spreading.\\

\noindent
A significant advance in this direction was the observation that the asymptotic behavior of the Lanczos coefficients is controlled by the high-frequency structure of the spectral measure associated with the initial operator. In particular, the authors of \cite{Parker:2018yvk} argued that exponential spectral tails imply asymptotically linear Lanczos growth, $b_n \sim \alpha n$, and proposed this behavior as a universal signature of chaotic quantum dynamics. This connection provides a remarkable bridge between operator growth and spectral analysis in that information about the long-time dynamics of Krylov complexity may be inferred directly from the asymptotic structure of the spectral function. The resulting picture has proven both physically compelling and surprisingly robust \cite{Avdoshkin:2022xuw, Hashimoto:2023swv, Erdmenger:2023wjg, Alishahiha:2024vbf, Baggioli:2024wbz}. A broad range of chaotic systems appear consistent with asymptotically linear Lanczos growth, while integrable or free systems typically exhibit slower asymptotic behavior\footnote{See also \cite{Bhattacharjee:2022vlt} which demonstrates that integrable systems exhibiting saddle-dominated scrambling may also exhibit linear growth.}. In this sense, the coefficient $\alpha$ has acquired a distinguished role as a dynamical measure of operator growth, analogous to the Lyapunov exponent in studies of out-of-time-order correlators \cite{Maldacena:2015waa}.\\

\noindent
At the same time, this raises a natural question. If the leading asymptotic behavior $b_n \sim \alpha n$ is fixed by the spectral tail, to what extent does this exhaust the asymptotic information contained in the Lanczos sequence? Put differently, are two systems with identical values of $\alpha$ necessarily asymptotically equivalent from the perspective of Krylov dynamics? The spectral-tail analysis underlying the classification of \cite{Parker:2018yvk} constrains only the behavior of the measure at large $|\omega|$, and is by construction insensitive to its global structure. Yet orthogonal-polynomial theory suggests that this global structure matters a great deal. Indeed, measures with identical tails but different \emph{topology}, for example, a support consisting of several disjoint bands rather than one, a gap that opens or closes as a parameter is varied, or an endpoint at which the density vanishes anomalously, possess qualitatively different recurrence asymptotics. From this perspective the coefficient $\alpha$ is only the first element of a much richer hierarchy, and the natural organizing principle for that hierarchy is not the tail of the spectral measure but its geometry.\\

\noindent
The central claim of this paper is that the topology of the spectral measure controls Krylov dynamics beyond the leading growth law, and that this control is both computable and physically consequential. A spectral gap renders the Lanczos coefficients quasiperiodic, with a frequency that is a topological invariant of the support. The closing of such a gap is not a smooth deformation but a genuine transition, at which the Lanczos sequence relaxes anomalously slowly and the crossover is governed by a Painlev\'e transcendent. Even within a fixed topological class, the local behavior of the measure at its endpoints leaves a definite imprint on the subleading Lanczos data. None of these phenomena are visible to tail-based arguments, and all of them are accessible once the problem is phrased in the language of orthogonal polynomials.\\ 

\noindent
The correspondence between Lanczos recursion and orthogonal polynomials is by now well established \cite{Muck:2022xfc} with the Krylov basis vectors being polynomial images of the seed operator, and the Lanczos coefficients the recurrence coefficients of the orthogonal-polynomial family associated with the spectral measure \cite{Caputa:2025ozd,Gamayun:2025hvu,Balasubramanian:2025xkj,Qu:2025lgo,Graef:2026pzv}. This identification makes available the asymptotic machinery of modern orthogonal-polynomial theory, and in particular its Riemann--Hilbert formulation \cite{deift2019riemannhilbertproblems}. In this context, the polynomials associated with a Krylov problem solve a matrix-valued Riemann--Hilbert problem of Fokas--Its--Kitaev type \cite{Fokas1992}, whose large-$z$ expansion encodes the Lanczos coefficients. Asymptotic questions about operator growth  become asymptotic questions about a Riemann--Hilbert problem, which the Deift--Zhou nonlinear steepest-descent method \cite{deift1993steepest} answers systematically. This strategy has recently been applied to Krylov dynamics by Lunt\footnote{We are grateful to Pawel Caputa for bringing this work to our attention.} \emph{et al.} \cite{r9v1-nxj1}, who use it to establish emergent random-matrix universality in the level-$n$ Green's function and to develop numerical methods for extracting hydrodynamic transport data. Our concerns in this article are complementary; where \cite{r9v1-nxj1} works within the single-cut setting and focuses on the interplay between high- and low-frequency behavior of the measure, we ask what happens when the support of the measure is topologically nontrivial.\\

\noindent
Concretely, we establish the following:
\begin{enumerate}
    \item For spectral measures with stretched-exponential tails $\rho(\omega) \sim e^{-c|\omega|^\beta}$, the Lanczos coefficients grow as $b_n \sim n^{1/\beta}$ with an explicit coefficient fixed by the equilibrium measure. The linear growth of the operator growth hypothesis \cite{Parker:2018yvk} is the $\beta = 1$ member of this Freud family, and the subleading corrections are determined by the local data of the measure at its endpoints.
    \item For gapped (multi-cut) spectra, the Lanczos coefficients oscillate quasiperiodically with a frequency fixed by the filling fraction of the spectral bands alone. The frequency is insensitive to every other feature of the measure, and we verify this prediction, in the Su--Schrieffer--Heeger chain and its next-nearest-neighbour deformation.
    \item At a gap-closing transition, two branch points of the spectral curve collide and its genus drops. The equilibrium density acquires a double zero at the merge point. The staggering of the Lanczos coefficients then decays anomalously slowly as $n^{-1/3}$, compared to the generic $n^{-2}$ of a single cut. The Hastings--McLeod solution of Painlev\'e~II interpolates between the gapped and merged phases, constituting a Krylov phase transition in the sense that we make precise in Section \ref{sec:critical-krylov}.
    \item In the conformal limit of the SYK model, the spectral measure is of Meixner--Pollaczek type and the Lanczos coefficients are exactly $b_n = \pi T \sqrt{n(n + 2/q - 1)}$. Consequently, while the operator scaling dimension $\Delta = 1/q$ leaves no trace in the leading growth rate $\alpha = \pi T$, we show that it is recovered from the constant offset $b_0 = \pi T(\Delta - \tfrac12)$.
\end{enumerate}

\noindent
Taken together, these results organize operator growth into a hierarchy where the tail exponent fixes the leading growth law, the topology of the support fixes the presence and frequency of oscillations, and the local structure of the measure fixes the subleading corrections. We also conjecture a cross-class law, \eqref{eq:prefactor-law}, for the universal $1/n$ correction to the Freud growth law, and verify it in three exactly solvable families.\\

\noindent
The remainder of the paper is organized as follows. Sections \ref{sec:orth-poly} and \ref{sec:RH-OP} recall the dictionary between Krylov dynamics and orthogonal polynomials and the Riemann--Hilbert formulation of the latter. Section \ref{sec:DZ} summarizes the Deift--Zhou steepest-descent analysis for single-cut measures, establishing the baseline against which the topological effects are measured, with the parametrix constructions deferred to Appendix \ref{app:steepest-descent}. Section \ref{sec:multi-cut} treats gapped spectra, where the theta-function parametrix produces quasiperiodic Lanczos coefficients, and tests the prediction in the SSH chain and its next-nearest-neighbour deformation. Section \ref{sec:critical-krylov} treats the gap-closing transition and its Painlev\'e~II crossover, with the local analysis in Appendix \ref{app:critical}. Section \ref{sec:refined-universality} assembles the results into a refined classification of operator growth and works out the application to the SYK model. Finally, we conclude in Section \ref{sec:discussion} with some open problems.\\

\section{Krylov dynamics and orthogonal polynomials}
\label{sec:orth-poly}

We begin by recalling the standard Krylov construction in a form that makes its relation to orthogonal polynomials manifest \cite{Muck:2022xfc}. Let \(\mathscr H_{\rm op}\) denote the Hilbert space of operators equipped with a positive inner product \((A|B)\), chosen so that the Liouvillian
\begin{eqnarray}
    \mathcal L=[H,\cdot]\,,
\end{eqnarray}
is self-adjoint \cite{Parker:2018yvk}. Given a normalized seed operator \(|\mathcal O_0)\), the Heisenberg evolution is given by
\begin{eqnarray}
    |\mathcal O(t))=
    e^{i\mathcal L t}|\mathcal O_0)\,.
\end{eqnarray}
The Lanczos procedure applied to the Krylov sequence
\begin{eqnarray}
    |\mathcal O_0),\quad 
    \mathcal L|\mathcal O_0),
    \quad \mathcal L^2|
    \mathcal O_0),\ldots,
\end{eqnarray}
produces an orthonormal basis \({|\mathcal O_n)}_{n\geq0}\) in which \(\mathcal L\) is tridiagonal,
\begin{eqnarray}
    \mathcal L|\mathcal O_n) 
    = b_{n+1}|\mathcal O_{n+1}) 
    + a_n|\mathcal O_n) 
    + b_n|\mathcal O_{n-1})
    \,, \qquad b_n>0.
\end{eqnarray}
For the operator-growth applications emphasized in much of the Krylov literature, the spectral measure is often even and the diagonal coefficients vanish, \(a_n=0\). We will retain the \(a_n\) in the present discussion, since it is natural from the orthogonal-polynomial point of view and will be useful later.
The autocorrelation function of the seed is
\begin{eqnarray}
    C(t) = 
    (\mathcal O_0|
    e^{i\mathcal L t}|
    \mathcal O_0)\,.
\end{eqnarray}
By the spectral theorem, there exists a positive measure \(\mu\) on \(\mathbb R\) such that
\begin{eqnarray}
    C(t)=\int_{\mathbb R}
    e^{i\omega t}\,
    d\mu(\omega).
\end{eqnarray}
Equivalently, the spectral function is
\begin{eqnarray}
    \Phi(\omega)=
    \int_{\mathbb R}
    e^{-i\omega t}C(t)\,dt\,,
\end{eqnarray}
whenever this Fourier transform exists in the usual or distributional sense. Consequently,
\[
    C(t)\longleftrightarrow
    \Phi(\omega)
    \longleftrightarrow
    d\mu(\omega).
\]
The measure \(\mu\) is the spectral measure of \(\mathcal L\) with respect to the cyclic vector \(|\mathcal O_0)\),
\begin{eqnarray}
    d\mu(\omega) =
    (\mathcal O_0|
    dE_{\mathcal L}
    (\omega)|\mathcal O_0)\,,
\end{eqnarray}
where \(E_{\mathcal L}\) is the projection-valued spectral measure of \(\mathcal L\).\\

\noindent
The moments of this measure are precisely the Liouvillian moments,
\begin{eqnarray}
    \mu_k = 
    \int_{\mathbb R}
    \omega^k\,d\mu(\omega) =
    (\mathcal O_0|
    \mathcal L^k|\mathcal O_0)\,.
\end{eqnarray}
Performing Gram--Schmidt orthonormalization on the monomials
\[
1,\omega,\omega^2,\ldots
\]
with respect to \(d\mu\) gives a family of orthogonal polynomials \({p_n(\omega)}\). These satisfy the famous three-term recurrence
\begin{eqnarray}
    \omega p_n(\omega) =
    b_{n+1}p_{n+1}(\omega)
    + a_np_n(\omega) +
    b_np_{n-1}(\omega),
\end{eqnarray}
with the same Jacobi coefficients \(a_n,b_n\) that appear in the Lanczos recursion. In particular, the Krylov vectors are polynomial images of the seed in the sense that
\begin{eqnarray}
    |\mathcal O_n) = 
    p_n(\mathcal L)|\mathcal O_0).
\end{eqnarray}
This then is the fundamental dictionary,
\[
C(t)
\longleftrightarrow
\Phi(\omega)
\longleftrightarrow
d\mu(\omega)
\longleftrightarrow
{p_n}
\longleftrightarrow
\{a_n,b_n\}.
\]
Each object contains the same information, but in a different representation. The correlation function \(C(t)\) emphasizes real-time dynamics; the spectral function \(\Phi(\omega)\) emphasizes high-frequency structure; the measure \(d\mu\) is the spectral-theoretic object; the polynomials \(p_n\) encode the cyclic representation of the Liouvillian; and the Lanczos coefficients \(\{a_n,b_n\}\) describe hopping on Krylov space.\\

\noindent
The Krylov wavefunction is obtained by expanding the Heisenberg-evolved operator in the Krylov basis as
\begin{eqnarray}
    |\mathcal O(t)) 
    = \sum_{n\geq0}
    \varphi_n(t)|\mathcal O_n)\,.
\end{eqnarray}
Using the polynomial representation,
\begin{eqnarray}
    \varphi_n(t) =
    \int_{\mathbb R}e^{i\omega
    t}
    \,p_n(\omega)\,d\mu(\omega)\,,
    \label{eq:int-rep}
\end{eqnarray}
up to conventional phase choices used in the Lanczos literature. The amplitudes themselves satisfy a discrete Schrödinger equation on the half-line,
\begin{eqnarray}
    i\partial_t\varphi_n
    = b_{n+1}\varphi_{n+1} +
    a_n\varphi_n
    + b_n\varphi_{n-1},
    \qquad \varphi_n(0)
    =\delta_{n0}\,.
\end{eqnarray}
The Krylov complexity of the original seed operator is then the first moment of this probability distribution,
\begin{eqnarray}
    K(t)=
    \sum_{n\geq0}n
    |\varphi_n(t)|^2\,.
\end{eqnarray}
As a result, the large-time behavior of \(K(t)\) is controlled by the large-\(n\) behavior of the Jacobi coefficients and by the large-\(n\), large-\(t\) asymptotics of the oscillatory integral defining \(\varphi_n(t)\).
This is the point at which orthogonal-polynomial methods enter naturally. The Lanczos problem is not merely analogous to the theory of orthogonal polynomials; it is in fact an instance of it. The Krylov chain is the Jacobi matrix of the measure \(d\mu\), and Krylov dynamics is the unitary evolution generated by this Jacobi matrix. In this sense, the spectral measure, the recurrence coefficients, and the Krylov wavefunctions are three equivalent ways of describing the same cyclic representation of the Liouvillian.\\

\noindent
Recent work has only sharpened this viewpoint. For example, in \cite{Muck:2022xfc} it was emphasized that the recursion method underlying Krylov complexity is precisely the recursion method of orthogonal-polynomial theory, with the Krylov basis generated by polynomials in the Liouvillian. Qu further showed, in the random-matrix setting, that the recursion coefficients of orthogonal polynomials admit a direct Krylov interpretation and coincide with averaged Lanczos data in the large-\(N\) continuum limit \cite{Qu:2025lgo}. These results motivate treating asymptotic questions in Krylov physics as asymptotic questions about orthogonal polynomials.\\

\noindent
The purpose of the present work is to exploit this observation systematically. Once the Krylov problem is written in orthogonal-polynomial language, the large-\(n\) asymptotics of \(b_n\), and eventually of \(\varphi_n(t)\) and \(K(t)\), can be studied by the Riemann--Hilbert steepest-descent method. The role of the RH formulation is not to replace the Lanczos construction, but to provide an analytic machinery for extracting the asymptotic structure of the same Jacobi data.

\section{Riemann--Hilbert formulation of orthogonal polynomials}
\label{sec:RH-OP}

The Krylov--orthogonal-polynomial dictionary of Section \ref{sec:orth-poly} converts the problem of operator growth into the problem of the large-$n$ asymptotics of a recurrence coefficient sequence. In this section we recall the analytic object that controls those asymptotics: the matrix-valued Riemann--Hilbert problem of Fokas--Its--Kitaev type \cite{Fokas1992, deift2019riemannhilbertproblems, deift1993steepest, deift1999orthogonal}. The essential point, equations \eqref{eq:bn} and \eqref{eq:an} below, is that the complete Jacobi data $\{a_n, b_n\}$ sits in the first subleading coefficient of the large-$z$ expansion of its solution. Everything in the remainder of the paper follows from asymptotic analysis of that expansion.\\

\noindent
We work in the standard setting of orthogonal polynomials on the real line. Let $d\mu(x) = w(x)\,dx$ be a positive measure on $\mathbb{R}$, with $w(x) > 0$ on its support and all moments finite,
\begin{equation}
    \int_{\mathbb{R}} |x|^{k}\, w(x)\, dx < \infty\,,
    \qquad k = 0,1,2,\ldots
\end{equation}
Let $\pi_n(x) = x^n + \text{lower powers}$ denote the monic orthogonal polynomial of degree $n$,
\begin{equation}
    \int_{\mathbb{R}} \pi_n(x)\,\pi_m(x)\, w(x)\, dx = h_n\,\delta_{nm}\,,
    \qquad h_n > 0\,,
\end{equation}
with associated orthonormal polynomials $p_n(x) = h_n^{-1/2}\pi_n(x)$. The monic polynomials satisfy
\begin{equation}
    x\,\pi_n(x) = \pi_{n+1}(x) + a_n \pi_n(x) + \beta_n \pi_{n-1}(x)\,,
    \quad \pi_{-1} = 0\,,\quad \pi_0 = 1\,,
    \quad \beta_n = \frac{h_n}{h_{n-1}}\,,
\end{equation}
and the orthonormal polynomials satisfy
\begin{equation}
    x\,p_n(x) = b_{n+1}p_{n+1}(x) + a_n p_n(x) + b_n p_{n-1}(x)\,,
    \quad b_n = \sqrt{\beta_n}\,.
    \label{eq:orthonormal-recurrence}
\end{equation}
The recurrence coefficients are therefore encoded in the squared norms $h_n$. The Riemann--Hilbert formulation packages $\pi_n$, its norm, and its Cauchy transform into a single matrix-valued analytic function, from which the $h_n$ can be read off. Next, we define the Cauchy transform of an integrable function $f$ by
\begin{equation}
    \mathcal{C}[f](z) = \frac{1}{2\pi i}\int_{\mathbb{R}}
    \frac{f(s)}{s - z}\, ds\,,
    \qquad z \in \mathbb{C}\setminus\mathbb{R}\,,
\end{equation}
whose non-tangential boundary values from the upper and lower half-planes satisfy the Sokhotski--Plemelj relation $\mathcal{C}_+[f] - \mathcal{C}_-[f] = f(x)$ for almost every $x \in \mathbb{R}$. For each $n \geq 1$ set
\begin{equation}
    Y_n(z) =
    \begin{pmatrix}
    \pi_n(z) & \mathcal{C}[\pi_n w](z)
    \\[0.8em]
    -2\pi i\, h_{n-1}^{-1}\pi_{n-1}(z) &
    -2\pi i\, h_{n-1}^{-1}\mathcal{C}[\pi_{n-1}w](z)
    \end{pmatrix}.
    \label{eq:Y-def}
\end{equation}
Then $Y_n$ is the unique solution of the following Riemann--Hilbert problem:
\begin{enumerate}
    \item $Y_n(z)$ is analytic for $z \in \mathbb{C}\setminus\mathbb{R}$;
    \item the boundary values on the real axis satisfy
    \begin{equation}
        Y_{n,+}(x) = Y_{n,-}(x)
        \begin{pmatrix} 1 & w(x) \\ 0 & 1 \end{pmatrix},
        \qquad x \in \mathbb{R}\,;
        \label{eq:Y-jump}
    \end{equation}
    \item as $z \to \infty$,
    \begin{equation}
        Y_n(z) = \left(\mathbb{I} + \frac{Y_{1,n}}{z}
        + \frac{Y_{2,n}}{z^2} + O(z^{-3})\right)
        z^{n\sigma_3}\,,
        \qquad
        z^{n\sigma_3} := \begin{pmatrix} z^n & 0 \\ 0 & z^{-n}\end{pmatrix}\,.
        \label{eq:Y-normalization}
    \end{equation}
\end{enumerate}
That \eqref{eq:Y-def} solves this problem follows from the Sokhotski--Plemelj relation together with the orthogonality of $\pi_n$ against all polynomials of lower degree. Uniqueness of this solution follows from the triangularity of the jump matrix. Both verifications are elementary and are collected in Appendix \ref{app:RH-verification}.\\

\noindent
The recurrence coefficients are obtained from the large-$z$ expansion \eqref{eq:Y-normalization}. Since $\pi_n$ is monic and orthogonal to all lower-degree polynomials, the Cauchy transforms in the second column of \eqref{eq:Y-def} have leading behaviour
\begin{equation}
    \mathcal{C}[\pi_n w](z) = -\frac{h_n}{2\pi i}\, z^{-n-1} + O(z^{-n-2})\,,
    \qquad
    \mathcal{C}[\pi_{n-1}w](z) = -\frac{h_{n-1}}{2\pi i}\, z^{-n} + O(z^{-n-1})\,,
    \label{eq:cauchy-leading}
\end{equation}
so that the off-diagonal entries of $Y_{1,n}$ are
\begin{equation}
    (Y_{1,n})_{12} = -\frac{h_n}{2\pi i}\,,
    \qquad
    (Y_{1,n})_{21} = -\frac{2\pi i}{h_{n-1}}\,.
\end{equation}
Their product is precisely the ratio of squared norms, giving
\begin{equation}
    \beta_n = (Y_{1,n})_{12}(Y_{1,n})_{21}\,,
    \qquad
    b_n = \sqrt{(Y_{1,n})_{12}(Y_{1,n})_{21}}\,.
    \label{eq:bn}
\end{equation}
The diagonal coefficient is encoded in the same expansion. Writing $\pi_n(z) = z^n + c_n z^{n-1} + O(z^{n-2})$, comparison of the three-term recurrence at order $z^n$ gives $a_n = c_n - c_{n+1}$, and since $c_n = (Y_{1,n})_{11}$,
\begin{equation}
    a_n = (Y_{1,n})_{11} - (Y_{1,n+1})_{11}\,.
    \label{eq:an}
\end{equation}

\noindent
Equations \eqref{eq:bn} and \eqref{eq:an} are the bridge on which the rest of this paper rests. Together they say that the Riemann--Hilbert problem encodes the complete Jacobi data of the orthogonal-polynomial system, and hence, via the dictionary of Section \ref{sec:orth-poly}, the complete Lanczos data of the Krylov problem. In other words, any method for controlling $Y_n(z)$ as $n \to \infty$ is therefore a method for controlling operator growth. The Deift--Zhou nonlinear steepest-descent method is just such a method, and we turn to it next.

\section{Steepest descent and the single-cut baseline}
\label{sec:DZ}

The Deift--Zhou nonlinear steepest-descent method \cite{deift1993steepest} extracts the large-$n$ behaviour of $Y_n$ directly from the weight $w(x)$, without explicit knowledge of the polynomials. It proceeds through a chain of explicit transformations
\begin{equation}
    Y_n \;\longrightarrow\; T_n \;\longrightarrow\; S_n
    \;\longrightarrow\; R_n\,,
    \label{eq:transformation-chain}
\end{equation}
each removing one source of oscillatory or exponentially growing behaviour, until what remains is a problem whose solution is $\mathbb{I}$ up to controlled corrections. In this section we summarise the construction for weights whose equilibrium measure is supported on a single interval $[A,B]$ and record the resulting baseline. The Lanczos coefficients converge to constants fixed by the band edges, with subleading corrections fixed by the local behaviour of the measure at those edges. The detailed constructions which include the resolvent computation of the equilibrium measure, the Airy and Bessel parametrices, and the small-norm analysis, are collected in Appendix \ref{app:steepest-descent}. Readers interested only in the topological phenomena of Sections \ref{sec:multi-cut} and \ref{sec:critical-krylov} may treat the single-cut case as a black box whose output is \eqref{eq:leading-asymptotics} and \eqref{eq:soft-corrections}.\\

\noindent
A comprehensive and rigorous treatment of the single-cut analysis in the Krylov context, including a careful discussion of the regularity conditions on the weight under which it is valid, has recently been given by Lunt \emph{et al.} \cite{r9v1-nxj1}; we refer the reader there and to \cite{deift1999orthogonal} for the technical hypotheses, which we assume throughout.

\subsection{Equilibrium measures, $g$-function, and the first transformation}
\label{subsec:gfunction}

It is convenient to work with weights in exponential form, $w(x) = e^{-nV(x)}$, with the $n$-dependence made explicit; $n$-independent weights can be recovered by a rescaling, as we do in Section \ref{sec:refined-universality}. The object that organises the large-$n$ limit is the equilibrium measure $\mu_{\mathrm{eq}}$, the unique minimiser of the logarithmic energy functional
\begin{equation}
    I[\nu] \;=\; -\iint \log|x - y|\, d\nu(x)\, d\nu(y)
    \;+\; \int V(x)\, d\nu(x)
    \label{eq:energy-functional}
\end{equation}
over probability measures $\nu$ on $\mathbb{R}$. Its relevance is that the zeros of $\pi_n$ distribute according to $\mu_{\mathrm{eq}}$ as $n \to \infty$, so $\mu_{\mathrm{eq}}$ governs the global structure of the polynomials at large degree. This is clear from the matrix-integral representation
\begin{equation}
    \pi_n(x) = \frac{\displaystyle\int \Big(\prod_{i=1}^{n} dx_i\Big)
    \prod_{i=1}^{n}(x - x_i)\;
    e^{-n\sum_i V(x_i)}\prod_{i<j}(x_i - x_j)^2}
    {\displaystyle\int \Big(\prod_{i=1}^{n} dx_i\Big)\;
    e^{-n\sum_i V(x_i)}\prod_{i<j}(x_i - x_j)^2}\,,
    \label{eq:matrix-integral}
\end{equation}
in which $\mu_{\mathrm{eq}}$ is the saddle point of the eigenvalue density; see \cite{deift1999orthogonal} for a rigorous statement. For now, we will assume that $\mu_{\mathrm{eq}}$ is \emph{single-cut}, supported on one interval $[A,B]$, and write $d\mu_{\mathrm{eq}}(x) = \rho_{\mathrm{eq}}(x)\,dx$. Varying \eqref{eq:energy-functional} gives the Euler--Lagrange equation
\begin{equation}
    V'(x) = 2\,\mathsf{P}\!\!\int_A^B dy\;
    \frac{\rho_{\mathrm{eq}}(y)}{x - y}\,,
    \label{eq:EL-density}
\end{equation}
where $\mathsf{P}$ denotes the principal value of the integral. Solving \eqref{eq:EL-density} by a standard resolvent argument (see Appendix \ref{app:equilibrium}) yields
\begin{equation}
    \rho_{\mathrm{eq}}(x) = \frac{1}{2\pi}\, h(x)\,
    \sqrt{(x-A)(B-x)}\,,
    \qquad
    h(x) = \frac{1}{2\pi i}\oint_{\Gamma}
    \frac{V'(s)\, ds}{(s-x)\sqrt{(s-A)(s-B)}}\,,
    \label{eq:vprimee}
\end{equation}
where $\Gamma$ is a contour encircling $[A,B]$ and the branch of the square root fixed by $\sqrt{(s-A)(s-B)} \sim s$ as $s \to \infty$. The function $h$ is analytic and strictly positive on a neighbourhood of $[A,B]$ in the regular case. The band edges themselves are fixed by the normalisation of $\mu_{\mathrm{eq}}$ and the absence of a $z^0$ term in the resolvent at infinity,
\begin{equation}
    \int_A^B \frac{V'(s)\, ds}{\sqrt{(s-A)(B-s)}} = 0\,,
    \qquad
    \int_A^B \frac{s\, V'(s)\, ds}{\sqrt{(s-A)(B-s)}} = 2\pi\,.
    \label{eq:endpoint-conditions}
\end{equation}
For the quadratic potential $V(x) = x^2/2$, for example, these give $A = -2$, $B = 2$ and $h(x) = 1$, so that $\mu_{\mathrm{eq}}$ is the Wigner semicircle.\\

\noindent
The $g$-function is the logarithmic potential of the equilibrium measure,
\begin{equation}
    g(z) = \int_A^B \log(z - x)\, d\mu_{\mathrm{eq}}(x)\,,
    \qquad z \in \mathbb{C}\setminus(-\infty, B]\,,
    \label{eq:g-def}
\end{equation}
with the logarithm cut along $(-\infty, B]$. Four properties drive the analysis:
\begin{enumerate}
    \item $g$ is analytic on $\mathbb{C}\setminus(-\infty,B]$;
    \item as $z \to \infty$, $g(z) = \log z - \sum_{k\geq1} m_k/(k z^k)$ with $m_k = \int x^k d\mu_{\mathrm{eq}}(x)$;
    \item the boundary values satisfy the Euler--Lagrange conditions
    \begin{align}
        g_+(x) + g_-(x) - V(x) &= \ell\,,
        \qquad x \in [A,B]\,,
        \label{eq:EL-equality}\\
        g_+(x) + g_-(x) - V(x) &< \ell\,,
        \qquad x \in \mathbb{R}\setminus[A,B]\,,
        \label{eq:EL-in-equality}
    \end{align}
    with $\ell$ the Robin constant;
    \item the jump of $g$ across the support is purely imaginary,
    \begin{equation}
        g_+(x) - g_-(x) = 2\pi i \int_x^B \rho_{\mathrm{eq}}(s)\, ds
        \;=:\; i\phi(x)\,,
        \qquad x \in [A,B]\,,
        \label{eq:phi-def}
    \end{equation}
    with $\phi$ real, positive and strictly decreasing on $(A,B)$, with $\phi(B) = 0$ and $\phi(A) = 2\pi$.
\end{enumerate}
Property 4 is the one that fails to have a single-valued counterpart in the multi-cut setting of Section \ref{sec:multi-cut}, and it is that failure that is the origin of every topological effect in this paper.\\

\noindent
The first transformation is then given by
\begin{equation}
    T_n(z) = e^{-n\ell\sigma_3/2}\, Y_n(z)\,
    e^{-n g(z)\sigma_3}\, e^{n\ell\sigma_3/2}\,,
    \qquad\mathrm{with}\,\,
    \sigma_3 = \begin{pmatrix} 1 & 0 \\ 0 & -1\end{pmatrix}\,.
    \label{eq:T-def}
\end{equation}
Since $e^{ng(z)} \sim z^n$ at infinity by property 2, the factor $e^{-ng(z)\sigma_3}$ cancels the divergent normalisation $z^{n\sigma_3}$ in \eqref{eq:Y-normalization}. A direct computation \cite{deift1999orthogonal} shows that $T_n$ is analytic off $\mathbb{R}$, satisfies $T_n(z) = \mathbb{I} + O(z^{-1})$ at infinity, and has jumps
\begin{align}
    T_{n,+}(x) &= T_{n,-}(x)
    \begin{pmatrix} e^{-in\phi(x)} & 1 \\ 0 & e^{in\phi(x)}\end{pmatrix},
    \qquad \mathrm{for}\,\,x \in (A,B)\,,
    \label{eq:T-jump-cut}\\
    T_{n,+}(x) &= T_{n,-}(x)
    \begin{pmatrix} 1 & e^{n(g_+ + g_- - V - \ell)} \\ 0 & 1\end{pmatrix},
    \qquad\mathrm{for}\,\, x \in \mathbb{R}\setminus[A,B]\,.
    \label{eq:T-jump-exterior}
\end{align}
By the strict inequality \eqref{eq:EL-in-equality} the exterior jump is $\mathbb{I} + O(e^{-cn})$ uniformly on compact subsets. The problem is now normalised at infinity, at the cost of the rapidly oscillating diagonal entries $e^{\pm i n \phi(x)}$ on the support. The Jacobi data remains encoded in the $z^{-1}$ coefficient through \eqref{eq:bn} and \eqref{eq:an}.

\subsection{Opening lenses}
\label{subsec:lenses}

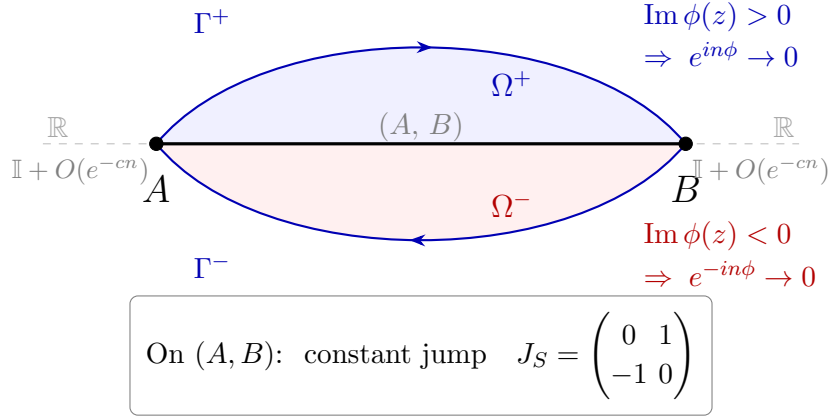
\begin{figure}[ht]
\centering
\begin{tikzpicture}[
    scale=1.0,
    >=Stealth,
    contour/.style={thick, blue!70!black,
        postaction={decorate,
            decoration={markings,
                mark=at position 0.52 with {\arrow{Stealth[length=6pt]}}
            }
        }
    },
    annot/.style={font=\small\itshape},
]
\def\Ax{-3.5}
\def\Bx{3.5}
\def\lensH{1.7}
\fill[blue!6]
    (\Ax,0) .. controls (\Ax+1.4,\lensH) and (\Bx-1.4,\lensH) .. (\Bx,0)
    -- cycle;
\fill[red!6]
    (\Ax,0) .. controls (\Ax+1.4,-\lensH) and (\Bx-1.4,-\lensH) .. (\Bx,0)
    -- cycle;
\draw[gray!50, dashed, thin] (-5.0,0) -- (5.0,0);
\draw[very thick] (\Ax,0) -- (\Bx,0);
\draw[contour]
    (\Ax,0) .. controls (\Ax+1.4,\lensH) and (\Bx-1.4,\lensH) .. (\Bx,0);
\draw[thick, blue!70!black,
    postaction={decorate,
        decoration={markings,
            mark=at position 0.52 with {\arrow{Stealth[length=6pt]}}
        }
    }]
    (\Bx,0) .. controls (\Bx-1.4,-\lensH) and (\Ax+1.4,-\lensH) .. (\Ax,0);
\filldraw[black] (\Ax,0) circle (2.5pt);
\filldraw[black] (\Bx,0) circle (2.5pt);
\node[below=8pt, font=\large\itshape] at (\Ax,0) {$A$};
\node[below=8pt, font=\large\itshape] at (\Bx,0) {$B$};
\node[blue!70!black, annot, above left=2pt]
    at (\Ax+1.2, 1.25) {$\Gamma^+$};
\node[blue!70!black, annot, below left=2pt]
    at (\Ax+1.2, -1.25) {$\Gamma^-$};
\node[blue!70!black, font=\normalsize] at (1.2, 0.8) {$\Omega^+$};
\node[red!70!black, font=\normalsize]  at (1.2,-0.8) {$\Omega^-$};
\node[gray, font=\small] at (0, 0.22) {$(A,\,B)$};
\node[blue!70!black, annot, align=left, anchor=west]
    at (2.8, 1.45)
    {$\mathrm{Im}\,\phi(z) > 0$\\[1pt]
     $\Rightarrow\; e^{in\phi} \to 0$};
\node[red!70!black, annot, align=left, anchor=west]
    at (2.8, -1.45)
    {$\mathrm{Im}\,\phi(z) < 0$\\[1pt]
     $\Rightarrow\; e^{-in\phi} \to 0$};
\node[gray, font=\footnotesize\itshape] at (-4.5, -0.35)
    {$\mathbb I + O(e^{-cn})$};
\node[gray, font=\footnotesize\itshape] at (4.5, -0.35)
    {$\mathbb I + O(e^{-cn})$};
\node[gray!70, font=\small] at (-4.8, 0.22) {$\mathbb{R}$};
\node[gray!70, font=\small] at (4.8, 0.22) {$\mathbb{R}$};
\node[draw=gray, rounded corners=3pt, inner sep=6pt,
      font=\small, align=center]
    at (0, -2.8)
    {On $(A,B)$:\; constant jump\quad
     $J_S = \begin{pmatrix} 0 & 1 \\ -1 & 0 \end{pmatrix}$};
\end{tikzpicture}
\caption{The lens-opening contour structure for the second transformation
$T_n \to S_n$. The interval $(A,B)$ (thick line) carries the constant jump
matrix $J_S$. The contours $\Gamma^\pm$ bound the lens regions $\Omega^\pm$,
in which the off-diagonal entries of the jump matrix decay exponentially as
$n \to \infty$ by \eqref{eq:phase-sign}. On $\mathbb{R}\setminus[A,B]$ the jump
is $\mathbb I + O(e^{-cn})$. The leading-order problem therefore reduces to a
constant-jump Riemann--Hilbert problem on $[A,B]$, solved by the global
parametrix $P^{(\infty)}$.}
\label{fig:lens-contours}
\end{figure}

The oscillatory entries $e^{\pm in\phi(x)}$ have no pointwise limit and are the main obstacle to any asymptotic analysis. They can be removed by the algebraic factorisation
\begin{equation}
    \begin{pmatrix} e^{-in\phi} & 1 \\ 0 & e^{in\phi}\end{pmatrix}
    =
    \begin{pmatrix} 1 & 0 \\ e^{in\phi} & 1 \end{pmatrix}
    \begin{pmatrix} 0 & 1 \\ -1 & 0 \end{pmatrix}
    \begin{pmatrix} 1 & 0 \\ e^{-in\phi} & 1 \end{pmatrix}\,,
    \label{eq:factorization}
\end{equation}
valid pointwise on $(A,B)$, and whose utility rests on the analytic continuation of the phase. Since $\rho_{\mathrm{eq}}$ is analytic and positive on $(A,B)$, $\phi$ extends analytically to a neighbourhood of the interval, and the Cauchy--Riemann equations give
\begin{equation}
    \mathrm{Im}\,\phi(z) > 0 \;\text{ for }\; \mathrm{Im}\, z > 0\,,
    \qquad
    \mathrm{Im}\,\phi(z) < 0 \;\text{ for }\; \mathrm{Im}\, z < 0\,,
    \label{eq:phase-sign}
\end{equation}
in a lens-shaped neighbourhood of $(A,B)$. Hence $e^{in\phi(z)}$ decays exponentially in the upper lens and $e^{-in\phi(z)}$ in the lower (Figure \ref{fig:lens-contours}).\\

\noindent
Fixing contours $\Gamma^{\pm}$ from $A$ to $B$ in the upper and lower half-planes respectively, bounding lens regions $\Omega^{\pm}$, we define
\begin{equation}
    S_n(z) = T_n(z) \times
    \begin{cases}
    \begin{pmatrix} 1 & 0 \\ -e^{in\phi(z)} & 1\end{pmatrix},
    & z \in \Omega^+,\\[3mm]
    \begin{pmatrix} 1 & 0 \\ e^{-in\phi(z)} & 1\end{pmatrix},
    & z \in \Omega^-,\\[3mm]
    \mathbb I, & \text{otherwise}\,.
    \end{cases}
    \label{eq:S-def}
\end{equation}
The resulting problem has a \emph{constant} jump $\left(\begin{smallmatrix}0&1\\-1&0\end{smallmatrix}\right)$ on $(A,B)$, and jumps $\mathbb{I} + O(e^{-cn})$ on $\Gamma^{\pm}$ and on $\mathbb{R}\setminus[A,B]$, uniformly on compact subsets away from the endpoints. All $n$-dependence is now exponentially suppressed except on the support itself, where it is absent.\\

\noindent
This already yields the leading asymptotic statement, and it is worth pausing on why. The surviving model problem depends on the weight \emph{only} through the two numbers $A$ and $B$. Every other feature of $w(x)$, from the shape of the density in the interior, to the higher moments, and even the detailed form of $V$, has been absorbed into the $g$-function and consequently into exponentially small corrections. The leading large-$n$ Lanczos asymptotics are therefore determined by the support of the equilibrium measure alone.

\subsection{The global parametrix and the leading asymptotics}
\label{subsec:global-parametrix}

The global parametrix $P^{(\infty)}(z)$ solves the model problem; It is analytic on $\mathbb{C}\setminus[A,B]$, with jump $\left(\begin{smallmatrix}0&1\\-1&0\end{smallmatrix}\right)$ on $(A,B)$ and $P^{(\infty)}(z) = \mathbb{I} + O(z^{-1})$ at infinity. Its solution is built from the conformal map of $\mathbb{C}\setminus[A,B]$ onto the exterior of the unit disc,
\begin{equation}
    \varphi(z) = \frac{2}{B-A}\left(z - \frac{A+B}{2}\right)
    + \frac{2}{B-A}\sqrt{\left(z - \frac{A+B}{2}\right)^2
    - \left(\frac{B-A}{2}\right)^2}\,,
    \label{eq:conformal-map}
\end{equation}
normalised so $\varphi(z) \sim 4z/(B-A)$ at infinity, together with the Szeg\H{o}-type function
\begin{equation}
    \gamma(z) = \left(\frac{z-B}{z-A}\right)^{1/4}\,,
    \label{eq:szego}
\end{equation}
analytic on $\mathbb{C}\setminus[A,B]$ with $\gamma(\infty) = 1$. In terms of these,
\begin{equation}
    P^{(\infty)}(z) =
    \begin{pmatrix}
    \dfrac{\gamma + \gamma^{-1}}{2} & \dfrac{\gamma - \gamma^{-1}}{2i}
    \\[3mm]
    -\dfrac{\gamma - \gamma^{-1}}{2i} & \dfrac{\gamma + \gamma^{-1}}{2}
    \end{pmatrix}\,,
    \label{eq:global-parametrix}
\end{equation}
where the jump follows from the boundary relation $\gamma_+(x) = i\gamma_-(x)$ on $(A,B)$ and the normalisation from $\gamma(\infty) = 1$. Expanding $\gamma(z) = 1 + (A-B)/4z + O(z^{-2})$ and applying \eqref{eq:bn} and \eqref{eq:an} along the chain $Y_n \to T_n \to S_n \approx P^{(\infty)}$ gives
\begin{equation}
    b_n \;\longrightarrow\; b_\infty = \frac{B-A}{4}\,,
    \qquad
    a_n \;\longrightarrow\; a_\infty = \frac{A+B}{2}\,,
    \qquad \mathrm{as}\,\,n \to \infty\,.
    \label{eq:leading-asymptotics}
\end{equation}
This is the single-cut baseline. For a weight with single-interval support and positive analytic density, the Lanczos coefficients converge to constants that are fixed entirely by the band edges, and with exponentially small corrections away from the endpoints. Each transformation play a distinct role; the $g$-function reduces the problem to the support, the lens opening renders everything but the constant jump exponentially negligible, and the parametrix solves what remaines in closed form.\\

\noindent
Two remarks deserve some attention at this point. First, \eqref{eq:leading-asymptotics} has a transparent dynamical interpretation. A Krylov chain with constant hopping $b_\infty$ and site energy $a_\infty$ is a translationally invariant tight-binding model with dispersion $\omega(k) = a_\infty + 2b_\infty\cos k$, whose band is exactly $[A,B]$. The Krylov chain reconstructs the spectral support from its own asymptotic band structure, and the operator wavefunction spreads ballistically along it. Second, under an affine rescaling $x \mapsto \lambda + \mu x$ of the spectral variable, the coefficients transform as $a_n \mapsto \lambda + \mu a_n$, $b_n \mapsto |\mu| b_n$, so \eqref{eq:leading-asymptotics} extends to any single-cut weight by translation and dilation. We will use this repeatedly in Section \ref{sec:refined-universality}, where the rescaling is $n$-dependent and converts a fixed weight with decaying tails into varying-weight form.

\subsection{Local parametrices and the subleading hierarchy}
\label{subsec:local-parametrices}

The global parametrix is, however, not uniformly valid near the endpoints. In particular, $\gamma(z)^{\pm1}$ develops fourth-root singularities as $z \to A, B$, and the exponential suppression of the lens jumps degenerates because $\phi$ vanishes there. Within discs $D_A, D_B$ of fixed small radius the problem must be solved exactly and matched to $P^{(\infty)}$ on the disc boundaries. Here, the finer structure of the weight enters and different endpoint behaviours produce different subleading corrections. Two cases matter for our purposes.\\

\noindent
\textbf{Soft edges:} For real-analytic $V$ with a regular equilibrium measure, generically the density vanishes as a square root,
\begin{equation}
    \rho_{\mathrm{eq}}(x) \sim c_B\sqrt{B-x}\,,
    \quad x \to B^-\,,
    \qquad
    c_B = \frac{h(B)}{2\pi}\sqrt{B-A} > 0\,,
    \label{eq:soft-edge}
\end{equation}
and similarly at $A$. The local parametrix is then built from Airy functions on the scale $\zeta \sim n^{2/3}(z-B)$; the construction is given in Appendix \ref{app:airy}. Comparing the exact solution to the assembled approximation via the final transformation $R_n$ and applying the small-norm theory (see Appendix \ref{app:small-norm}) gives
\begin{equation}
    b_n = \frac{B-A}{4}\left(1 + \frac{c_b}{n^2} + O(n^{-3})\right),
    \qquad
    a_n = \frac{A+B}{2} + \frac{c_a}{n^2} + O(n^{-3})\,,
    \label{eq:soft-corrections}
\end{equation}
with explicit constants $c_a, c_b$ built from $h(A)$, $h(B)$ and the endpoint conformal data. Here too, two features deserve emphasis. The corrections begin at order $n^{-2}$ since the $O(n^{-1})$ contributions from the two endpoints cancel in the recurrence coefficients, as manifest by the small-norm expansion. Second, the $c_a, c_b$ depend on the weight only through the local data $h(A), h(B)$, so two weights sharing a support and its endpoint densities share subleading asymptotics through this order.\\

\noindent
\textbf{Hard edges:} A qualitatively different local behaviour arises when the weight carries an algebraic singularity at an endpoint that survives the large-$n$ limit,
\begin{equation}
    w(x) = (x-A)^{\alpha}\, e^{-nV(x)}\,, \qquad \mathrm{for}\,\,\alpha > -1\,.
    \label{eq:hard-edge-weight}
\end{equation}
In this case $\rho_{\mathrm{eq}}(x) \sim \tilde c_A (x-A)^{-1/2}$ diverges as an inverse square root rather than vanishing. The Airy parametrix at $A$ is then replaced by a Bessel parametrix built from $I_{\alpha}, K_{\alpha}$ on the scale $\tilde\zeta \sim n^2(z-A)$ (see Appendix \ref{app:bessel} for the detailed construction), and the $n^{-2}$ coefficient acquires an explicit $\alpha$-dependence. For the classical Jacobi weights $w(x) = (x-A)^{\alpha}(B-x)^{\alpha'}$ the result can be checked against the exactly known Jacobi recurrence coefficients,
\begin{equation}
    b_n = \frac{B-A}{4}\left(1 + \frac{c(\alpha,\alpha')}{n^2}
    + O(n^{-3})\right),
    \label{eq:jacobi-check}
\end{equation}
and the agreement is exact.\\

\noindent
The single-cut analysis therefore organises the large-$n$ Lanczos data into a three-level hierarchy:
\begin{enumerate}
    \item the \emph{leading constants} $a_\infty = (A+B)/2$, $b_\infty = (B-A)/4$ are fixed by the support of the equilibrium measure alone;
    \item the \emph{rate of approach and the correction coefficients} are fixed by the local structure of the weight at the endpoints; specifically the values $h(A), h(B)$ at a soft edge, and the exponents $\alpha, \alpha'$ at hard edges;
    \item \emph{exponentially small} corrections encode the remaining global data of the weight.
\end{enumerate}
Consequently, two weights sharing a support share leading asymptotics but can be distinguished by their correction series whenever their endpoint data differ. This is a key point of the refined classification that we propose in Section \ref{sec:refined-universality}.

\subsection{Decaying tails and the Hermite class}
\label{subsec:gaussian}

The analysis so far assumed a compactly supported equilibrium measure. Spectral measures in Krylov dynamics are typically supported on all of $\mathbb{R}$ with decaying tails, and the bridge between the two settings is the $n$-dependent rescaling anticipated at the end of Section \ref{subsec:global-parametrix}. We illustrate it on the case where the answer is known exactly, the Gaussian measure
\begin{equation}
    d\mu(\omega) = \frac{1}{\sqrt{2\pi}}\, e^{-\omega^2/2}\, d\omega\,,
    \label{eq:gaussian-measure}
\end{equation}
which arises in Krylov dynamics whenever the spectral density of a many-body seed is controlled by a central limit theorem, as for the finite-density probe states of \cite{Graef:2026pzv}. The associated orthogonal polynomials are precisely the Hermite polynomials $\mathsf{H}_n$, with exactly known recurrence coefficients
\begin{equation}
    b_n = \sqrt{n}\,,\qquad a_n = 0\,,\qquad n \geq 1\,.
    \label{eq:hermite-exact}
\end{equation}
The $\sqrt{n}$ growth defines the \emph{Hermite class}. Our purpose here is to see how it emerges from the steepest-descent analysis, since the same mechanism produces the general Freud law of Section \ref{sec:refined-universality}.
The weight \eqref{eq:gaussian-measure} has no endpoints, so our analysis in Section \ref{subsec:global-parametrix} does not apply to it directly. Passing to varying-weight form $w_n(\omega) = e^{-n\omega^2/2}$, the degree-$n$ monic polynomial with respect to $w_n$ is related to the standard Hermite polynomial by
\begin{equation}
    \pi_n(\omega) = n^{-n/2}\,\mathsf{H}_n(\sqrt{n}\,\omega)\,,
    \label{eq:hermite-rescaling}
\end{equation}
so the fixed-weight problem at degree $n$ is equivalent to the varying-weight problem at degree $n$ viewed on the scale $\omega \sim \sqrt{n}$. The potential is now $V(\omega) = \omega^2/2$, for which \eqref{eq:endpoint-conditions} gives $A = -2$, $B = 2$ and $h \equiv 1$; the equilibrium measure is the Wigner semicircle
\begin{equation}
    d\mu_{\mathrm{eq}}(\omega) = \frac{1}{2\pi}\sqrt{4-\omega^2}\; d\omega\,,
    \qquad \omega \in [-2,2]\,,
    \label{eq:semicircle}
\end{equation}
with $g$-function
\begin{equation}
    g(z) = \frac{1}{2}\Big(z\sqrt{z^2-4}
    - 2\log\big(z+\sqrt{z^2-4}\big) + \log 4 + 1\Big)\,,
    \label{eq:gaussian-g}
\end{equation}
cut on $[-2,2]$. The chain $Y_n \to T_n \to S_n$ runs exactly as in Sections \ref{subsec:gfunction}--\ref{subsec:lenses}, and the global parametrix is \eqref{eq:global-parametrix} with $\gamma(z) = \big((z-2)/(z+2)\big)^{1/4}$.\\

\noindent
The one key difference from the compactly supported case is the exterior jump. Since $w_n$ is nonzero on all of $\mathbb{R}$, \eqref{eq:T-jump-exterior} is not vacuous outside $[-2,2]$; its exponent is
\begin{equation}
    g_+(\omega)+g_-(\omega)-\frac{\omega^2}{2}-\ell
    = -\frac{1}{2}\Big(|\omega|\sqrt{\omega^2-4}
    - 2\log\big(|\omega|+\sqrt{\omega^2-4}\big) + \log 4\Big)\,,
    \label{eq:gaussian-EL-exterior}
\end{equation}
which is strictly negative for $|\omega|>2$ and behaves as $-\omega^2/4$ for large $|\omega|$. The exterior jump is therefore  \emph{super-exponentially} small in $n$, and the analysis proceeds unchanged.\\

\noindent
At $\omega = \pm2$ the equilibrium density vanishes as a square root, so the local parametrices are the Airy ones of Section \ref{subsec:local-parametrices}. Here, however, the expansion terminates and all subleading corrections in the $R_n$ expansion \eqref{eq:R-expansion} vanish identically,\footnote{The exactness of $b_n = \sqrt{n}$ is a consequence of the ladder structure $x = a + a^{\dagger}$ with $[a,a^{\dagger}]=1$, which is the algebraic origin of the Hermite three-term recurrence; equivalently, the Gaussian is the unique weight for which multiplication by $x$ factorises into first-order raising and lowering operators. From the Riemann--Hilbert side this appears as an exact cancellation; for $V(x)=x^2/2$ the local parametrix matching at the endpoints produces no residual error, and the asymptotic expansion terminates at leading order.} giving
\begin{equation}
    b_n^{(\mathrm{rescaled})} = 1 \qquad\text{exactly.}
    \label{eq:rescaled-bn-exact}
\end{equation}
Undoing the rescaling $\omega = x/\sqrt{n}$ recovers $b_n = \sqrt{n}$, in agreement with \eqref{eq:hermite-exact}. What the Riemann--Hilbert route adds is a decomposition of that formula. The factor $1$ is the global parametrix, fixed by the semicircular equilibrium measure on $[-2,2]$, while the factor $\sqrt{n}$ is entirely the rescaling that absorbed the $n$-dependence of the varying weight. Section \ref{sec:refined-universality} generalises exactly this split, with the tail exponent controlling the rescaling and the equilibrium measure controlling the prefactor.\\

\noindent
In applications, the spectral measure is only approximately Gaussian with deviations controlled by a large parameter $L$ such as the system size, chain length, or the number of degrees of freedom. Suppose then that
\begin{equation}
    d\mu_L(\omega) = \frac{1}{\sqrt{2\pi}}e^{-\omega^2/2}
    \left(1 + \frac{\delta\rho_1(\omega)}{\sqrt{L}}
    + \frac{\delta\rho_2(\omega)}{L} + \cdots\right) d\omega\,,
    \label{eq:near-gaussian}
\end{equation}
which is the structure predicted by the quantum central limit theorem for finite-density states with short-range correlations with the $\delta\rho_k$ being determined by the cumulants of the Hamiltonian. The perturbation enters the steepest-descent analysis multiplicatively in the jump matrix and therefore modifies the $R_n$ problem relative to the Gaussian baseline. Tracing it through the chain gives
\begin{equation}
    b_n = \sqrt{n}\left(1 + \frac{r_1(n)}{L} + O(L^{-2})\right),
    \label{eq:near-gaussian-bn}
\end{equation}
where $r_1$ is a bounded function of $n$ fixed by the third and fourth cumulants. The $O(L^{-1/2})$ term is absent by parity since an odd cumulant produces an antisymmetric perturbation of the weight. This shifts the diagonal coefficients $a_n$ but not the off-diagonal $b_n$. The leading correction to $b_n$ is therefore $O(L^{-1})$ and is controlled by the excess kurtosis of the energy distribution. The growth exponent is consequently exactly $1/2$ for any measure in the domain of attraction of the Gaussian, with deviations entering only as multiplicative corrections vanishing in the thermodynamic limit. The Hermite class is stable.\\

\noindent
The Lanczos growth $b_n = \sqrt{n}$, $a_n = 0$ gives the Krylov wavefunction equation
\begin{equation}
    i\partial_t\phi_n = \sqrt{n+1}\,\phi_{n+1} + \sqrt{n}\,\phi_{n-1}\,,
    \label{eq:hermite-schrodinger}
\end{equation}
which, under $\phi_n \leftrightarrow \langle n|\psi(t)\rangle$ with $|n\rangle$ the number basis of a bosonic mode, is generated by the displacement Hamiltonian $H = a + a^{\dagger}$. Since $[a,a^{\dagger}]$ is central the evolution factorises,
$e^{-it(a+a^{\dagger})} = e^{-t^2/2}e^{-ita^{\dagger}}e^{-ita}$, and the seed $\phi_n(0)=\delta_{n0}$ evolves into the coherent state $|{-it}\rangle$. The Krylov distribution is therefore Poisson, $|\phi_n(t)|^2 = e^{-\bar n(t)}\bar n(t)^n/n!$, with
\begin{equation}
    K(t) = \bar n(t) = t^2
    \label{eq:hermite-krylov-complexity}
\end{equation}
exactly, for all $t$. The absence of corrections is the dynamical counterpart of the exactness of \eqref{eq:hermite-exact}; both follow from the ladder structure. At finite $L$ the $r_1(n)/L$ term in \eqref{eq:near-gaussian-bn} deforms that algebra, and on any fixed time interval $K(t) = t^2(1+O(L^{-1}))$ so the quadratic law is robust, and only its exact coefficient and the vanishing of subleading terms are properties of the Gaussian point. Quadratic growth places the Hermite class strictly between the integrable class of Section \ref{subsec:global-parametrix}, where bounded $b_n$ give ballistic $K(t)\sim t$, and the chaotic class, where $b_n \sim \alpha n$ gives $K(t)\sim e^{2\alpha t}$. Exponential complexity growth requires spectral tails decaying no faster than exponentially; the Gaussian tail is too thin, and $\sqrt{n}$ Lanczos growth produces only polynomial spreading in Krylov space. This ordering is the content of the Freud law, to which we return in Section \ref{sec:refined-universality}.

\section{Gapped spectra and quasiperiodic Krylov dynamics}
\label{sec:multi-cut}

The analysis of Section \ref{sec:DZ} presupposed a single-interval support, and that assumption entered at exactly one place, property 4 of the $g$-function, equation \eqref{eq:phi-def}, which produced a single-valued phase $\phi$ across the whole support. When the support has several components, however, that construction fails. It leaves behind an $n$-dependent oscillatory phase which no steepest-descent transformation can remove, because it is a topological invariant of the support rather than a feature of the weight. The Lanczos coefficients subsequently inherit that phase and become quasiperiodic.\\

\noindent
Physically, a disconnected spectral measure corresponds to a seed operator whose spectral density has a gap \textit{i.e.} a range of Liouvillian eigenvalues that the seed does not probe. Such spectra are ubiquitous. Essentially any gapped band structure produces one and they sit entirely outside the tail-based classification, which constrains only the behaviour of the measure at large $|\omega|$ and is blind to whether the support is connected. They also sit outside the single-cut universality established in \cite{r9v1-nxj1}, whose authors note that their semicircle law for the level-$n$ Green's function could be violated locally if the spectral function contained spectral gaps. This section works out what happens when it does.

\subsection{Two-cut equilibrium measures and the elliptic curve}
\label{subsec:two-cut-eq}

Toward this end, consider a spectral measure supported on two intervals,
\begin{equation}
    \operatorname{supp}(\mu) = [\omega_1^-,\omega_1^+]
    \cup [\omega_2^-,\omega_2^+]\,,
    \qquad \omega_1^+ < \omega_2^-\,,
    \label{eq:two-cut-support}
\end{equation}
with absolutely continuous density positive on the interior of each band, and gap $(\omega_1^+,\omega_2^-)$.
Two distinct problems lead to this configuration, and it is worth separating them at the outset because they arise in different applications and are governed by different, albeit ultimately equivalent, objects.\\

\noindent
\textbf{(i) Varying weights:} For $w = e^{-nV}$ with a potential whose equilibrium measure is two-cut, the density is
\begin{equation}
    \rho_{\mathrm{eq}}(\omega) = \frac{1}{2\pi}\,
    \frac{|P(\omega)|}{\sqrt{|Q(\omega)|}}\, h(\omega)\,,
    \qquad
    Q(\omega) = \prod_{j=1,2}(\omega-\omega_j^-)(\omega-\omega_j^+)\,,
    \label{eq:two-cut-density}
\end{equation}
with $P$ a polynomial fixed by $V$ subject to nonnegativity of $\rho_{\mathrm{eq}}$ on the support. This is the setting of double-scaling limits in matrix models.\\

\noindent
\textbf{(ii) Fixed measures in the Szeg\H{o} class:} For an $n$-independent measure on a compact set $E$, as is the case, for example, in a lattice model with a gapped band structure, there is no potential at all. The governing object is instead the equilibrium measure of the \emph{set} $E$, that is the $V\equiv0$ minimiser of \eqref{eq:energy-functional} constrained to $E$, or equivalently the harmonic measure of $E$ at infinity. For two intervals this is
\begin{equation}
    d\omega_E(\omega) = \frac{|\omega - c|}
    {\pi\sqrt{|Q(\omega)|}}\, d\omega\,,
    \label{eq:harmonic-measure}
\end{equation}
where the constant $c$ lies in the gap and is fixed by normalisation, equivalently by
\begin{equation}
    \int_{\omega_1^+}^{\omega_2^-}
    \frac{(\omega - c)\, d\omega}{\sqrt{|Q(\omega)|}} = 0\,.
    \label{eq:c-condition}
\end{equation}
For measures in the Szeg\H{o} class on several intervals the recurrence coefficients are asymptotically almost periodic \cite{Widom1969}.\\

\noindent
The two settings differ in what determines the density on the support, but they share the feature that matters here. In both, the analytic structure is controlled by
\begin{equation}
    \mathcal{R}\;:\; y^2 = Q(z)\,,
    \label{eq:elliptic-curve}
\end{equation}
a hyperelliptic curve with four branch points or equivalently, an elliptic curve, of genus one. The single-cut case had $Q$ quadratic and $\mathcal{R}$ of genus zero. This then is the main difference from which everything else follows; the underlying complex geometry is a torus rather than a sphere.

\subsection{The irremovable phase}
\label{subsec:irremovable-phase}

The transformations $Y_n \to T_n \to S_n$ proceed as in Sections \ref{subsec:gfunction}--\ref{subsec:lenses}. The $g$-function normalises the problem at infinity, and lenses opened around each band reduce the jumps on $\Gamma^{\pm}$ to $\mathbb{I}+O(e^{-cn})$. After the second transformation, $S_n$ carries:
\begin{enumerate}
    \item constant jumps $\left(\begin{smallmatrix}0&1\\-1&0\end{smallmatrix}\right)$ on each band $(\omega_j^-,\omega_j^+)$;
    \item a jump on the \emph{gap},
    \begin{equation}
        S_{n,+}(\omega) = S_{n,-}(\omega)
        \begin{pmatrix} e^{-2\pi i n\Omega} & 0 \\
        0 & e^{2\pi i n\Omega}\end{pmatrix},
        \qquad \omega \in (\omega_1^+,\omega_2^-)\,,
        \label{eq:gap-jump}
    \end{equation}
    where
    \begin{equation}
        \Omega = \int_{\omega_1^-}^{\omega_1^+}
        d\mu_{\mathrm{eq}}(\omega) \;\in\; (0,1)
        \label{eq:filling-fraction}
    \end{equation}
    is the mass the equilibrium measure assigns to the first band. The two
    settings of Section \ref{subsec:two-cut-eq} differ here in a way that
    matters for what follows. In setting (ii) $\Omega$ is determined by the
    support alone, whereas in
    setting (i) it additionally depends on $V$ through the equilibrium
    problem;
    \item jumps $\mathbb{I}+O(e^{-cn})$ on the lens boundaries and on $\mathbb{R}\setminus\operatorname{supp}(\mu)$.
\end{enumerate}

\noindent
Item 2 has no counterpart in the single-cut analysis and is the crux of this section. It arises because $g_+ - g_-$ is constant but \emph{nonzero} across the gap. The phase \eqref{eq:phi-def} accumulated across the first band is $2\pi\Omega$ rather than the full $2\pi$, and the deficit must be carried somewhere. Unlike every other jump surviving the second transformation, \eqref{eq:gap-jump} is not exponentially small; it is a pure phase of modulus one, oscillating in $n$. No further deformation removes it, because $\Omega$ is determined by the equilibrium measure and hence by the support, and the support is, by construction, respected by the transformations.\\

\noindent
The model problem after the second transformation therefore has constant jumps on two disjoint intervals \emph{and} an oscillatory diagonal jump on the gap. Its solution is no longer elementary. to see this, let $\tau$ be the modular parameter of $\mathcal{R}$, the ratio of periods of the holomorphic differential $dz/y$,
\begin{equation}
    \tau = \frac{\oint_{\mathfrak b} d\omega/\sqrt{Q(\omega)}}
    {\oint_{\mathfrak a} d\omega/\sqrt{Q(\omega)}}\,,
    \qquad \operatorname{Im}\tau > 0\,,
    \label{eq:modular-parameter}
\end{equation}
with $\mathfrak a$ encircling the first band and $\mathfrak b$ connecting the two, and let
\begin{equation}
    \vartheta(u\,|\,\tau) = \sum_{m\in\mathbb{Z}}
    e^{\pi i m^2\tau + 2\pi i m u}
    \label{eq:theta-function}
\end{equation}
be the associated Riemann theta function. The global parametrix is built from ratios of theta functions with shifted arguments,
\begin{equation}
    P^{(\infty)}(z) = M(z)\,
    \frac{\vartheta\big(\mathcal{A}(z) - n\Omega - d\,\big|\,\tau\big)}
    {\vartheta\big(\mathcal{A}(z) - d\,\big|\,\tau\big)}\,,
    \label{eq:two-cut-parametrix-schematic}
\end{equation}
schematically, with $\mathcal{A}(z) = \int_{\omega_1^-}^{z} d\omega/\sqrt{Q(\omega)}$ the Abel map, $d$ fixed by normalisation and $M$ collecting prefactors. The full matrix structure involves theta functions with characteristics and is given in \cite{10.1155/S1073792897000500}. The essential point is that the parametrix depends on $n$ only through $n\Omega$, and only modulo the period lattice, because $\vartheta$ is quasiperiodic, as can be seen already in \eqref{eq:two-cut-parametrix-schematic}.

\subsection{Quasiperiodic Lanczos coefficients}
\label{subsec:quasiperiodic}

Extracting the recurrence coefficients from the $z^{-1}$ coefficient of \eqref{eq:two-cut-parametrix-schematic} through \eqref{eq:bn} and \eqref{eq:an} gives the central result of this paper,
\begin{equation}
    b_n = \bar b + \tilde b\,\cos(2\pi n\Omega + \varphi_0)
    + O(e^{-cn})\,,
    \quad
    a_n = \bar a + \tilde a\,\cos(2\pi n\Omega + \varphi_0')
    + O(e^{-cn})\,.
    \label{eq:oscillatory-bn}
\end{equation}
The Lanczos sequence does not converge. It oscillates about a mean at frequency $\Omega$, with exponentially small corrections, and since $\Omega$ is generically irrational the oscillation is quasiperiodic rather than periodic.\\

\noindent
Let us be precise about which data in \eqref{eq:oscillatory-bn} is topological and which is not, since this is the distinction the numerics of Section \ref{subsec:solvable} test.
\begin{itemize}
    \item The \emph{frequency} $\Omega$ is fixed by the support alone,
    through \eqref{eq:filling-fraction}; here and throughout the
    applications below we are in setting (ii), where this holds
    unconditionally. In particular, it does not depend on the density of the measure on that support.
    \item The \emph{mean, amplitude and phase} $\bar b, \tilde b$, and
        $\varphi_0$ are not
        topological. 
        They involve the
        Szeg\H{o} function of the density and therefore depend on the shape of the measure, not only on its support. In the special case where the measure \emph{is} the equilibrium measure they reduce to explicit functions of the modular parameter $\tau$, and hence of the band edges, but in general they do not.
\end{itemize}
This asymmetry means that the frequency can be computed from the four band edges alone, in advance and without reference to the dynamics, while the remaining three parameters cannot. Note that, since $n$ is an integer,
\begin{equation}
    \cos(2\pi n\Omega + \varphi_0)
    = \cos\big(2\pi n(1-\Omega) - \varphi_0\big)\,,
    \label{eq:aliasing}
\end{equation}
so $\Omega$ and $1-\Omega$ are  \emph{exactly indistinguishable} in the Lanczos sequence. This is the statement that the two bands enter symmetrically in the sense that the sequence knows the partition of the support into bands but not which band was labelled first. Any frequency extracted from data is therefore determined only up to $\Omega \mapsto 1-\Omega$.\\

\noindent
Dynamically, a quasiperiodic Lanczos sequence means the Krylov chain is no longer asymptotically translation invariant. It carries an incommensurate modulation of its hoppings, which partially reflects the Krylov wavefunction and produces a spreading pattern richer than the ballistic propagation of the single-cut case. The complexity acquires a beating envelope,
\begin{equation}
    K(t) \sim 2\bar b\, t + \mathcal{F}(t)\,,
    \label{eq:beating-complexity}
\end{equation}
with $\mathcal{F}$ bounded and quasiperiodic with amplitude controlled by $\tilde b/\bar b$. A wide gap gives pronounced beating and periodic refocusing of the wavefunction near the origin of the chain; a narrow gap gives exponentially suppressed beating and dynamics approaching the single-cut ballistic regime.
The extension to $p+1$ bands is immediate and requires only replacing the elliptic curve by a genus-$p$ hyperelliptic one. The global parametrix involves the genus-$p$ theta function $\vartheta(\boldsymbol{u}\,|\,\boldsymbol{\tau})$ with $\boldsymbol{u}\in\mathbb{C}^p$ and $\boldsymbol{\tau}$ the $p\times p$ period matrix, and the Lanczos coefficients carry $p$ independent frequencies, one per gap,
\begin{equation}
    b_n = \bar b + \sum_{j=1}^{p}\tilde b_j
    \cos(2\pi n\Omega_j + \varphi_j) + O(e^{-cn})\,,
    \qquad
    \Omega_j = \int_{\text{$j$-th band}} d\mu_{\mathrm{eq}}\,.
    \label{eq:multi-cut-bn}
\end{equation}
The number of independent frequencies in the asymptotic Lanczos data is thus given by the genus of the spectral curve, which in turn is the number of spectral gaps. This is the sharpest form of our claim that Krylov dynamics detects spectral topology.

\subsection{Solvable realisations: dimerised chains}
\label{subsec:solvable}

The prediction \eqref{eq:oscillatory-bn} is worth testing in a setting where the spectral measure is known in closed form and the band edges can be read off analytically. Gapped free-fermion chains provide the minimal such setting and we treat two here: the Su--Schrieffer--Heeger (SSH) chain, whose chiral symmetry forces the degenerate rational point $\Omega=\tfrac12$ and which serves as a warm-up, and a next-nearest-neighbour deformation of the SSH chain which breaks the chiral symmetry and realises the generic quasiperiodic regime.

\subsubsection*{The symmetric case: SSH}

The SSH model \cite{Su_Schrieffer_Heeger_1979} is a tight-binding chain with two sites per unit cell and alternating hoppings $t_1$ (intracell), $t_2$ (intercell),
\begin{equation}
    H = \sum_j \big( t_1\, c^{\dagger}_{j,A}c_{j,B}
    + t_2\, c^{\dagger}_{j+1,A}c_{j,B} + \mathrm{h.c.}\big)\,.
    \label{eq:ssh-H}
\end{equation}
We will take as seed a single local Majorana operator $\chi_0$. Since $H$ is quadratic, $\mathcal{L}=[H,\cdot]$ preserves the single-fermion sector and acts there with the single-particle spectrum of $H$, so the spectral measure is the local density of states of the Bloch Hamiltonian. With
\begin{equation}
    \varepsilon(k) = \pm\sqrt{t_1^2+t_2^2+2t_1t_2\cos k}\,,
    \qquad k\in[-\pi,\pi]\,,
    \label{eq:ssh-disp}
\end{equation}
the support is
\begin{equation}
    \operatorname{supp}(\mu) = [-\omega_{\max},-\Delta]
    \cup [\Delta,\omega_{\max}]\,,
    \qquad \Delta = |t_1-t_2|\,,\quad \omega_{\max}=t_1+t_2\,,
    \label{eq:ssh-support}
\end{equation}
which is \eqref{eq:two-cut-support} with a gap of width $2\Delta$ centred at the origin, closing at the dimerisation-symmetric point $t_1=t_2$. Evaluating the density of states gives the measure in closed form,
\begin{equation}
    d\mu(\omega) = \frac{|\omega|\, d\omega}
    {\pi\sqrt{(\omega_{\max}^2-\omega^2)(\omega^2-\Delta^2)}}\,,
    \qquad \Delta<|\omega|<\omega_{\max}\,,
    \label{eq:ssh-measure}
\end{equation}
with inverse-square-root divergences at all four band edges.\footnote{These are the one-dimensional van Hove singularities. A van Hove singularity is a non-smooth feature in the density of states at an energy where the band dispersion is flat, $\nabla_k\varepsilon(k)=0$; in one dimension the band extrema produce inverse-square-root divergences $\rho(\omega)\sim|\omega-\omega_{\rm edge}|^{-1/2}$. For \eqref{eq:ssh-disp} the flat points are $k=0,\pi$, which map to $\pm\Delta$ and $\pm\omega_{\max}$, and it is there that the factor $[(\omega_{\max}^2-\omega^2)(\omega^2-\Delta^2)]^{-1/2}$ diverges. See \cite{ashcroft_solid_1976}.} These are hard edges in the sense of Section \ref{subsec:local-parametrices}, resolved by Bessel rather than Airy parametrices.\\

\begin{figure}[t]
  \centering
  \begin{subfigure}[b]{0.49\textwidth}
    \centering
    \includegraphics[width=\textwidth]{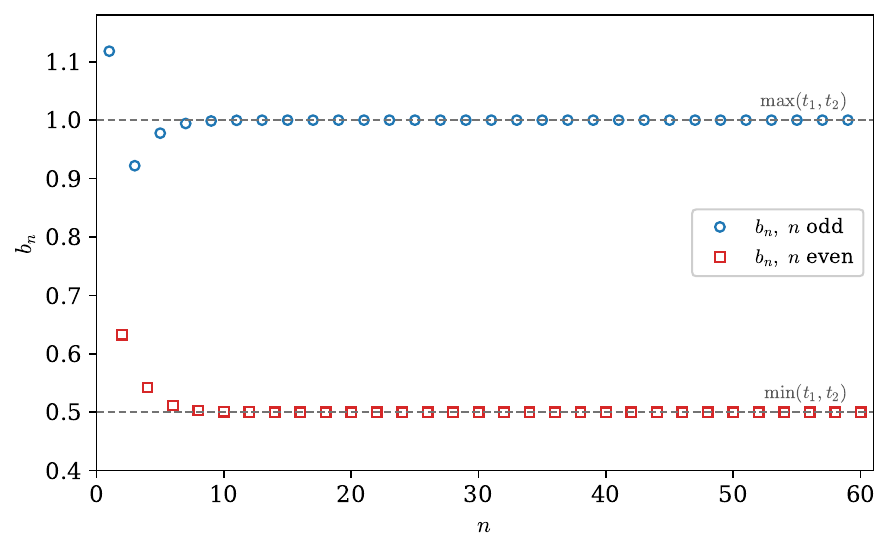}
    \caption{}\label{fig:ssh-continuum}
  \end{subfigure}
  \hfill
  \begin{subfigure}[b]{0.49\textwidth}
    \centering
    \includegraphics[width=\textwidth]{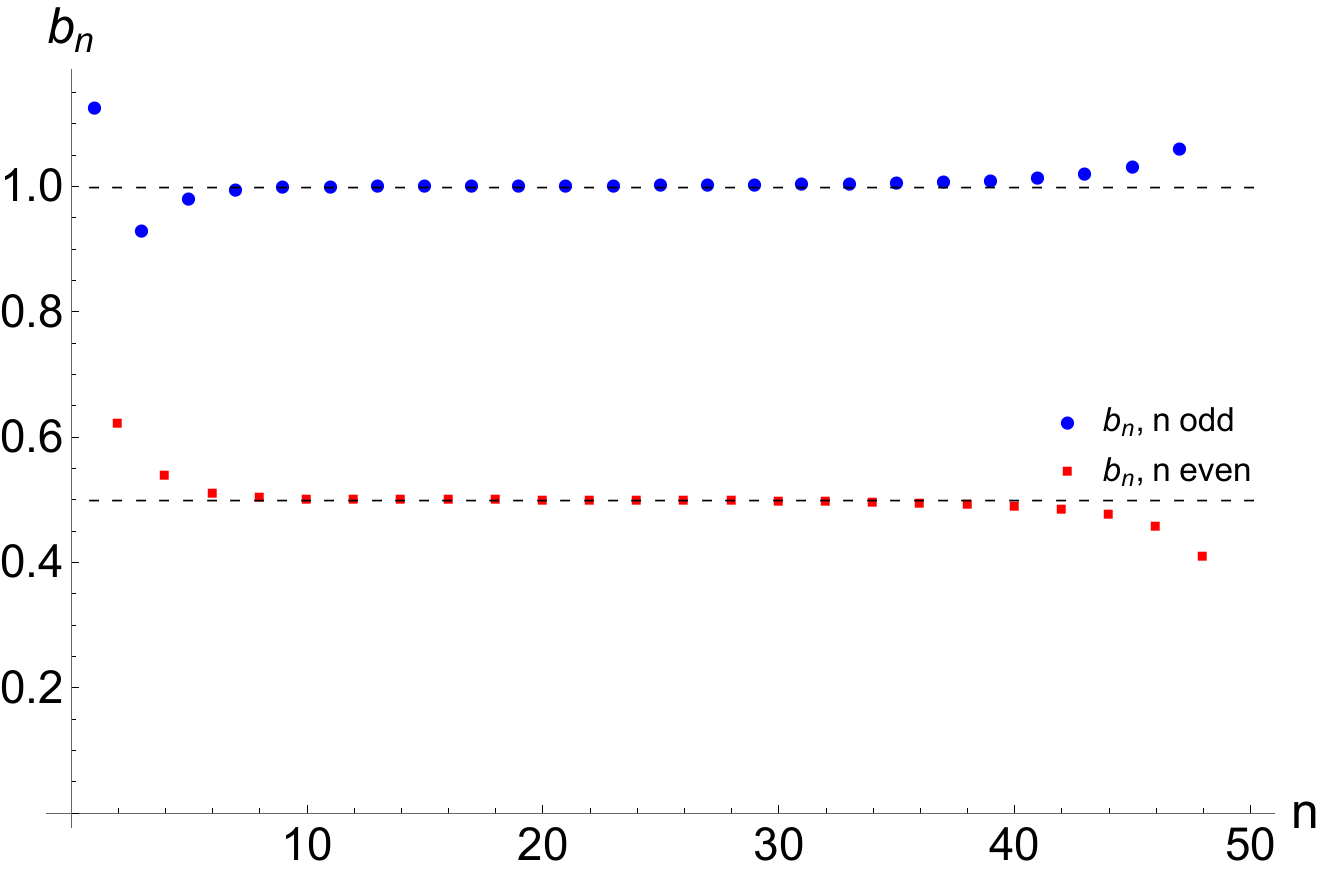}
    \caption{}\label{fig:ssh-discrete}
  \end{subfigure}
  \caption{Even--odd staggering of the Lanczos coefficients for the SSH
    chain with $t_1=1$, $t_2=\tfrac12$, whose spectral measure
    \eqref{eq:ssh-measure} is supported on the symmetric two-cut set
    $[-\tfrac32,-\tfrac12]\cup[\tfrac12,\tfrac32]$. The sequence splits into
    two branches converging to $\max(t_1,t_2)$ and $\min(t_1,t_2)$ (dashed),
    the exact limits \eqref{eq:ssh-exact-limits} predicted by the
    $\Omega=\tfrac12$ degeneration \eqref{eq:ssh-stagger} of the two-cut
    formula; the diagonal coefficients vanish identically,
    $\max_n|a_n| = 4\times10^{-16}$.
    (\subref{fig:ssh-continuum}) Coefficients computed from the moments of the
    continuum measure \eqref{eq:ssh-measure}, agreeing with
    \eqref{eq:ssh-exact-limits} to machine precision.
    (\subref{fig:ssh-discrete}) The same quantity from the Lanczos algorithm
    applied to the momentum band discretised into $50$ points, with the
    infinite-temperature thermal state as reference.}
  \label{fig:ssh}
\end{figure}

\noindent
The measure is even, so $a_n=0$ identically, and the reflection $\omega\to-\omega$ forces $\Omega=\tfrac12$. At that value \eqref{eq:oscillatory-bn} degenerates and
\begin{equation}
    b_n = \bar b + \tilde b\cos(\pi n + \varphi_0) + O(e^{-cn})
    = \bar b \pm \tilde b\cos\varphi_0 + O(e^{-cn})\,,
    \label{eq:ssh-stagger}
\end{equation}
so the sequence splits into two exponentially converging branches for even and odd $n$. This even--odd staggering is the behaviour long associated with a spectral gap in the Krylov literature; here it appears as the $\Omega=\tfrac12$ member of the two-cut family. The limits can be obtained exactly. The substitution $x=\omega^2$ maps \eqref{eq:ssh-measure} to the arcsine measure
$d\nu(x) \propto dx/\sqrt{(\omega_{\max}^2-x)(x-\Delta^2)}$ on $[\Delta^2,\omega_{\max}^2]$, whose monic recurrence coefficients are the Chebyshev values $\alpha_k = (\omega_{\max}^2+\Delta^2)/2$ and $\beta_k = (\omega_{\max}^2-\Delta^2)^2/16$ for $k\geq2$. The standard symmetrisation relations $\alpha_k = b_{2k}^2+b_{2k+1}^2$, $\beta_k = b_{2k-1}^2b_{2k}^2$ then give, writing $p,q$ for the even- and odd-index limits,
\begin{equation}
    p^2+q^2 = \frac{\omega_{\max}^2+\Delta^2}{2}\,,
    \qquad
    pq = \frac{\omega_{\max}^2-\Delta^2}{4}\,,
\end{equation}
from which we write $p+q = \omega_{\max}$ and $|p-q| = \Delta$, so that
\begin{equation}
    \{p,q\} = \{\max(t_1,t_2),\ \min(t_1,t_2)\}\,,
    \qquad
    \bar b = \frac{t_1+t_2}{2}\,,
    \qquad
    \tilde b = \frac{|t_1-t_2|}{2}\,.
    \label{eq:ssh-exact-limits}
\end{equation}
The staggering amplitude is half the gap width, vanishing as $\Delta\to0$ and returning $b_n\to\omega_{\max}/2$, the single-cut result for the uniform chain. Numerically the approach to \eqref{eq:ssh-exact-limits} is exponential at rate $(\min/\max)^n$, consistent with the $O(e^{-cn})$ error in \eqref{eq:ssh-stagger}; Figure \ref{fig:ssh} shows the comparison, with $a_n$ vanishing to machine precision.\\

\noindent
Two caveats about this example deserve special mention. First, $\Omega=\tfrac12$ and the even--odd staggering both follow from the reflection symmetry alone, so the agreement in Figure \ref{fig:ssh} confirms that the machinery reproduces a known answer without probing the genus-one structure specifically. Adding a staggered on-site potential (the Rice--Mele deformation \cite{RiceMele1982}) widens the gap but preserves $\omega\to-\omega$ and hence leaves $\Omega=\tfrac12$ unchanged. Second, and less obviously, the quadratic map shows that the SSH two-cut structure is a double cover of a single-cut problem. In the variable $x=\omega^2$ the measure is supported on one interval for every $\Delta$, including $\Delta=0$. We return to the consequences of this in the following section.

\subsubsection*{The generic case: breaking chiral symmetry}

To reach the quasiperiodic regime of \eqref{eq:oscillatory-bn}, with irrational $\Omega$ and $a_n\neq0$, the two bands must be made inequivalent. The minimal way to do this within the same model is to retain the SSH hoppings and add a same-sublattice next-nearest-neighbour hopping $t'$, equal on both sublattices, giving the Bloch Hamiltonian \cite{AsbothOroszlanyPalyi2016}
\begin{equation}
    H(k) = \begin{pmatrix} 2t'\cos k & h(k) \\
    h(k)^{*} & 2t'\cos k\end{pmatrix},
    \qquad h(k) = t_1 + t_2 e^{-ik}\,,
    \label{eq:nnn-H}
\end{equation}
with bands
\begin{equation}
    \varepsilon_{\pm}(k) = 2t'\cos k
    \pm\sqrt{t_1^2+t_2^2+2t_1t_2\cos k}\,.
    \label{eq:nnn-disp}
\end{equation}
The added term is proportional to the identity, so it shifts both bands by the same $k$-dependent amount without affecting the eigenvectors. Two consequences follow immediately. The sublattice weight of every Bloch state remains $\tfrac12$, so the local spectral measure of a single-site seed is still the normalised density of states of \eqref{eq:nnn-disp}. And the additive $2t'\cos k$ breaks the chiral symmetry $\varepsilon\to-\varepsilon$. Consequently, the bands are no longer mirror images, the measure is no longer even, $a_n\neq0$ and $\Omega\neq\tfrac12$.\\

\noindent
For the measure to remain a genuine two-cut density with only the four band-edge singularities, $t'$ must be weak enough that neither band develops an interior extremum. Differentiating \eqref{eq:nnn-disp} gives critical points at $k=0,\pi$ together with any solution of $\sqrt{f(k)} = t_1t_2/(2t')$ inside the band, where $f(k)=t_1^2+t_2^2+2t_1t_2\cos k$. Since $\sqrt{f}$ ranges over $[|t_1-t_2|, t_1+t_2]$, an interior extremum appears precisely for
$t_1t_2/2(t_1+t_2) < t' < t_1t_2/2|t_1-t_2|$, so taking
\begin{equation}
    t' < \frac{t_1t_2}{2(t_1+t_2)}
    \label{eq:nnn-cond}
\end{equation}
keeps both bands monotone in $k$ with edges at $k=0,\pi$. These are again hard edges resolved by Bessel parametrices. We take $t_1=1$, $t_2=\tfrac35$, $t'=\tfrac{3}{20}$, comfortably below the threshold $t_1t_2/2(t_1+t_2)=\tfrac{3}{16}$. Evaluating \eqref{eq:nnn-disp} at $k=0,\pi$ gives the two asymmetric bands
\begin{equation}
    \operatorname{supp}(\mu) = [-1.3,\,-0.7]\cup[0.1,\,1.9]\,.
    \label{eq:nnn-bands}
\end{equation}

\begin{figure}[t]
  \centering
  \centering
  \begin{subfigure}[b]{0.49\textwidth}
    \centering
    \includegraphics[width=\textwidth]{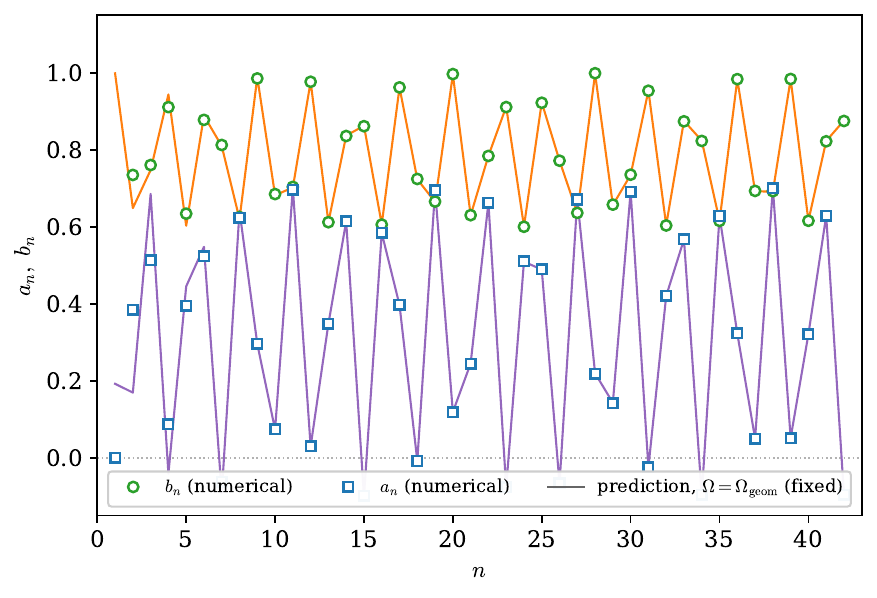}
    \caption{}\label{fig:ssh-continuum}
  \end{subfigure}
  \hfill
  \begin{subfigure}[b]{0.49\textwidth}
    \centering
    \includegraphics[width=\textwidth]{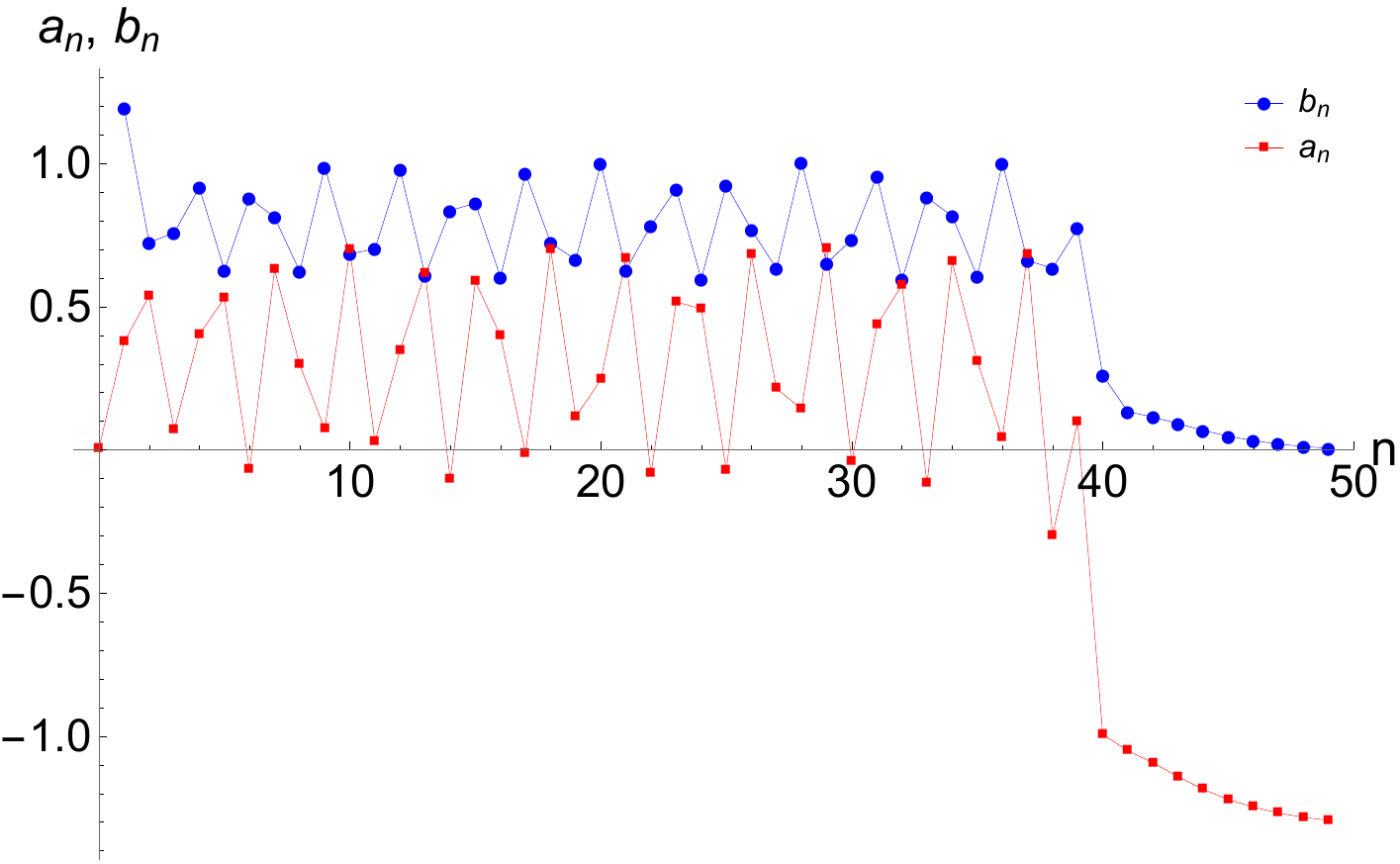}
    \caption{}\label{fig:ssh-discrete}
  \end{subfigure}
  
  \caption{Quasiperiodic Lanczos coefficients for the chain \eqref{eq:nnn-H}
    with $t_1=1$, $t_2=\tfrac35$, $t'=\tfrac{3}{20}$, whose spectral measure is
    supported on the asymmetric bands \eqref{eq:nnn-bands}. Circles and squares
    are the numerical $b_n$ and $a_n$; curves are the two-cut prediction
    \eqref{eq:oscillatory-bn} with the oscillation frequency \emph{fixed,
    without fitting}, to the harmonic measure $\Omega = 0.3700$ of the support
    computed from the band edges through \eqref{eq:harmonic-measure}. Only
    means, amplitudes and phases are fit. Both sequences oscillate at this
    common geometric frequency even though the physical density differs entirely
    from the equilibrium measure for which \eqref{eq:oscillatory-bn} was derived.
    (a) from the moments of the continuum measure. (b) from
    the Lanczos algorithm on the momentum band discretised into $50$ points at
    infinite temperature; agreement persists for the first $\sim40$
    coefficients, beyond which the finite-$L$ discrete spectrum departs from
    the continuum one, the range of agreement growing with $L$.}
  \label{fig:twocut}
\end{figure}

\noindent
Unlike the chirally symmetric case the measure has no elementary closed form, and its recurrence coefficients must be computed numerically from its moments. The frequency, however, does not require the measure at all. Solving \eqref{eq:c-condition} for the bands \eqref{eq:nnn-bands} gives $c = -0.3237$, and integrating \eqref{eq:harmonic-measure} over the lower band gives
\begin{equation}
    \Omega = 0.3700\,,
    \qquad 1-\Omega = 0.6300\,,
    \label{eq:nnn-omega}
\end{equation}
computed from the four band edges alone, with no input from the dynamics and no fitting.
Figure \ref{fig:twocut} compares this against the numerically computed Lanczos coefficients. Both $b_n$ and $a_n$ are quasiperiodic and both track \eqref{eq:oscillatory-bn} with the frequency \emph{fixed} to \eqref{eq:nnn-omega}; only the means, amplitudes and phases are fit, as they must be, since  those are not determined by the support.\\

\noindent
The force of the comparison is that the physical density of \eqref{eq:nnn-disp} bears no resemblance to the equilibrium measure for which the theta-function asymptotics were derived. It is not in the same family, and its shape on the support is entirely different. Yet the oscillation frequency is the same, to four digits, because the frequency is a property of the support and nothing else. Physically, the next-nearest-neighbour hopping tilts the two bands asymmetrically and the Krylov chain inherits an incommensurate modulation of its hoppings and site energies, rather than the exact period-two dimerisation of the pure SSH chain. This quasiperiodic Krylov chain whose modulation frequency encodes the filling fraction of the spectral bands, is the generic operator-growth signature of a gapped, non-chirally-symmetric spectrum. In addition, its Krylov complexity exhibits a beating envelope \eqref{eq:beating-complexity} at an irrational rather than commensurate beat.

\section{Gap closing and Krylov phase transitions}
\label{sec:critical-krylov}

The single- and two-cut regimes of Sections \ref{sec:DZ} and \ref{sec:multi-cut} are separated by a boundary at which a spectral gap closes and two bands merge. Crossing it changes the genus of the spectral curve, and with it the qualitative character of the Lanczos sequence, with quasiperiodic oscillation on one side and monotone convergence on the other. In this section we describe the crossover, which turns out to be governed by a Painlev\'e transcendent that exhibits an anomalously slow relaxation, sharply distinguishing it from both phases.

\subsection{Branch-point collision and the degeneration of the curve}
\label{subsec:branch-collision}

In the two-cut phase the equilibrium density involves $\sqrt{Q}$ with
$Q(\omega) = \prod_{j}(\omega-\omega_j^-)(\omega-\omega_j^+)$, and vanishes as a square root at each of the four endpoints. Now let the inner endpoints approach one another, $\omega_1^+ \to \omega_c \leftarrow \omega_2^-$. Then two of the four branch points of $\sqrt{Q}$ then collide and the quartic degenerates,
\begin{equation}
    Q(z) \;\longrightarrow\; (z-\omega_1^-)(z-\omega_c)^2(z-\omega_2^+)\,,
    \label{eq:degenerate-quartic}
\end{equation}
so that the square root factorises as
\begin{equation}
    \sqrt{Q(z)} = (z-\omega_c)\sqrt{(z-\omega_1^-)(z-\omega_2^+)}\,.
    \label{eq:factorized-root}
\end{equation}
The factor $(z-\omega_c)$ is rational and carries no branch cut. The curve $\mathcal{R}: y^2=Q(z)$, of genus one in the two-cut phase, degenerates to genus zero where the torus pinches to a sphere, the $\mathfrak a$-cycle collapses, and the modular parameter $\tau\to i\infty$. By \eqref{eq:two-cut-parametrix-schematic} the theta functions degenerate to exponentials and the oscillatory structure of \eqref{eq:oscillatory-bn} disappears while the amplitude $\tilde b$ vanishing as the gap narrows.\\

\noindent
The signature of this on the merged support is a non-generic zero of the density. To fix its order, note that on $[\omega_1^-,\omega_2^+]$ the density retains the single-cut form \eqref{eq:vprimee},
\begin{equation}
    \rho_{\mathrm{eq}}(\omega) = \frac{1}{2\pi}\,h(\omega)\,
    \sqrt{(\omega-\omega_1^-)(\omega_2^+-\omega)}\,,
    \label{eq:critical-density}
\end{equation}
in which the square-root factor is bounded away from zero at $\omega_c$. Any zero of $\rho_{\mathrm{eq}}$ at an interior point must therefore be a zero of $h$. However, since $\rho_{\mathrm{eq}}\geq0$ on its support, any interior zero of $h$ cannot be of odd order as the density would change sign across it. The generic possibility is a double zero, and it is exactly the development of such a zero that marks the transition,
\begin{equation}
    \rho_{\mathrm{eq}}(\omega) \sim \kappa\,(\omega-\omega_c)^2\,,
    \qquad \omega\to\omega_c\,,
    \qquad
    \kappa = \frac{h''(\omega_c)}{4\pi}
    \sqrt{(\omega_c-\omega_1^-)(\omega_2^+-\omega_c)} > 0\,.
    \label{eq:quadratic-vanishing}
\end{equation}
This is distinct from both the square-root vanishing at the outer endpoints as well as from the strictly positive density in the interior of a generic single-cut support. This quadratic zero is the defining local signature of the transition.\\

\noindent
The same conclusion follows from the two-cut side by a degree count, which also identifies where the two powers of $(\omega-\omega_c)$ come from. Writing the resolvent as $W = \tfrac12\big(V' - M\sqrt{Q}\big)$ with polynomial $M$, we find that $\deg M = d-2$ in the one-cut phase and $\deg M = d-3$ in the two-cut phase, where $d = \deg V$. At the transition $Q_4 = (z-\omega_c)^2 Q_2$, and matching the two representations gives
\begin{equation}
    h_{\text{1-cut}}(z) = (z-\omega_c)\, h_{\text{2-cut}}(z)\,.
    \label{eq:h-degeneration}
\end{equation}
One power of $(\omega-\omega_c)$ is supplied by the collapsing square root \eqref{eq:factorized-root} and one by the vanishing of $h_{\text{2-cut}}$ at the merge point. For the symmetric quartic potential, where $h_{\text{2-cut}}(z) = gz$ and $\omega_c=0$, this gives $\rho_{\mathrm{eq}}\to\frac{g}{2\pi}\,\omega^2\sqrt{b^2-\omega^2}$ directly. Since only the order of the zero was used in our argument, the exponent must be universal for the gap-closing transition, independent of the potential or of the details of the bands.

\subsection{The Painlev\'e II parametrix}
\label{subsec:painleve}

The Airy parametrix of Section \ref{subsec:local-parametrices} is built from solutions of $y''=\zeta y$. It is adapted to a square-root zero of the density and hence to a $3/2$-power zero of the phase. At $\omega_c$ the density vanishes quadratically, so writing $u = \omega-\omega_c$ and using \eqref{eq:phi-def}, we find that
\begin{equation}
    \phi(u) - \phi(0) = -2\pi\!\int_0^u\!\rho_{\mathrm{eq}}
    \;\sim\; -\frac{2\pi\kappa}{3}\,u^3\,,
    \qquad u\to0\,.
    \label{eq:cubic-phase}
\end{equation}
As in Section \ref{subsec:local-parametrices} the exact problem is solved inside a disc $D_c$ of fixed small radius $\delta$ centred at $\omega_c$ and matched to the global parametrix on its boundary $\partial D_c$; the difference is that the disc must now accommodate a critical region of width $n^{-1/3}$ rather than $n^{-2/3}$. The model problem inside $D_c$ is solved by the $\Psi$-function of the \emph{Painlev\'e II equation},
\begin{equation}
    q''(s) = s\,q(s) + 2q(s)^3\,,
    \label{eq:painleve-II}
\end{equation}
with Hastings--McLeod boundary condition $q(s)\to\mathrm{Ai}(s)$ as $s\to+\infty$. This is the unique real solution positive on all of $\mathbb{R}$, decaying exponentially at $+\infty$ and satisfying
\begin{equation}
    q(s) \sim \sqrt{-s/2}\,,
    \label{eq:HM-minus-infinity}
\end{equation}
as $s\to-\infty$. The $\Psi$-function solves the associated Lax pair,
\begin{equation}
    \frac{\partial\Psi}{\partial\zeta} =
    \begin{pmatrix}
    -4i\zeta^2 - i(s+2q^2) & 4\zeta q + 2iq' \\[1mm]
    4\zeta q - 2iq' & 4i\zeta^2 + i(s+2q^2)
    \end{pmatrix}\Psi\,,
    \label{eq:PII-lax}
\end{equation}
and carries the phase $\theta(\zeta,s) = \tfrac43\zeta^3 + s\zeta$. Pulling this all together, the critical parametrix is then given by
\begin{equation}
    P^{(\mathrm{crit})}(z) = E^{(\mathrm{crit})}_n(z)\,
    \Psi\big(\zeta_n(z), s_n\big)\,
    e^{i\theta(\zeta_n,s_n)\sigma_3}\,,
    \label{eq:PII-parametrix}
\end{equation}
where $E^{(\mathrm{crit})}_n$ is fixed by matching to $P^{(\infty)}$ on $\partial D_c$.\\

\noindent
The local variable and the deformation variable are both fixed by requiring $n\phi$ and $\theta$ to agree term by term. Deforming the potential so that
$\rho_{\mathrm{eq}}\sim\kappa\,(u^2+\sigma)$ near the merge point, \eqref{eq:cubic-phase} becomes
$\phi(u)-\phi(0) = -2\pi\kappa\big(\tfrac13 u^3 + \sigma u\big)$, and matching the cubic and linear terms separately gives
\begin{equation}
    \zeta_n = c_1\,n^{1/3}u\,,
    \qquad
    s_n = c_0\, n^{2/3}\sigma\,,
    \qquad
    c_1 = \Big(\frac{\pi\kappa}{2}\Big)^{1/3},
    \qquad
    c_0 = 2^{4/3}(\pi\kappa)^{2/3}\,.
    \label{eq:local-variables}
\end{equation}
Both constants are explicit in terms of the single local datum $\kappa$ of \eqref{eq:quadratic-vanishing}. The critical region $|u|\lesssim n^{-1/3}$ is parametrically wider than the $n^{-2/3}$ Airy region at a generic endpoint, and this is the origin of the parametrically larger corrections obtained below.
Note that in the two-cut phase $\sigma<0$ and $\rho_{\mathrm{eq}}$ vanishes on $|u|<\sqrt{|\sigma|}$, so
\begin{equation}
    \epsilon_{\mathrm{gap}} \;=\; \omega_2^- - \omega_1^+ \;=\; 2\sqrt{|\sigma|}\,,
    \qquad\text{equivalently}\qquad
    \sigma = -\tfrac14\,\epsilon_{\mathrm{gap}}^{\,2}\,.
    \label{eq:sigma-gap}
\end{equation}
In other words, the variable that scales as $n^{2/3}$ is the deformation parameter of the potential, while the gap width enters quadratically. The transition is therefore resolved on the scale $\epsilon_{\mathrm{gap}}\sim n^{-1/3}$.

\subsection{Double scaling and the critical exponent}
\label{subsec:double-scaling}

The double-scaling limit takes $\sigma\to0$ and $n\to\infty$ with $s = c_0 n^{2/3}\sigma$ held fixed, so that $s<0$ in the two-cut phase, $s=0$ at the transition, and $s>0$ in the merged phase. The size of the correction follows from the order of the matching mismatch. In contrast to the Airy case, where the expansion proceeds in powers of $\zeta^{-3/2}$, the large-$\zeta$ expansion of the Painlev\'e $\Psi$-function is organised in powers of $\zeta^{-1}$,
\begin{equation}
    \Psi(\zeta,s)\,e^{i\theta(\zeta,s)\sigma_3}
    = \mathbb I + \frac{\Psi_1(s)}{\zeta} + O(\zeta^{-2})\,,
    \qquad
    (\Psi_1)_{12},\,(\Psi_1)_{21} \propto q(s)\,.
    \label{eq:PII-large-zeta}
\end{equation}
On $\partial D_c$ we have $|\zeta| = c_1 n^{1/3}\delta$ by \eqref{eq:local-variables}, so the mismatch with the global parametrix is
\begin{equation}
    \Delta_n = P^{(\mathrm{crit})}\big(P^{(\infty)}\big)^{-1} - \mathbb I
    \;=\; \frac{E^{(\mathrm{crit})}_n \Psi_1 \big(E^{(\mathrm{crit})}_n\big)^{-1}}
    {c_1 n^{1/3}\delta} + O(n^{-2/3})
    \;=\; O(n^{-1/3})\,,
    \label{eq:critical-mismatch}
\end{equation}
which tends to zero as the small-norm theory requires. Its Cauchy integral over $\partial D_c$ is fixed by the residue at $\omega_c$ and is therefore $\delta$-independent,
\begin{equation}
    R^{(1)} = \frac{1}{2\pi i}\oint_{\partial D_c}
    \frac{\Delta_n(\varsigma)}{\varsigma - z}\, d\varsigma
    \;=\; O(n^{-1/3})\,.
    \label{eq:critical-R1}
\end{equation}
Unwinding the chain of transformations through \eqref{eq:bn} then gives corrections of the same order in the recurrence coefficients. Since the off-diagonal entries of $\Psi_1$ carry $q(s)$, and  $\Omega\to\tfrac12$ at the transition, the correction alternates in sign with $n$, giving
\begin{equation}
    b_n = b_\infty\left(1 + \frac{(-1)^n\, d_1\, q(s)}{n^{1/3}}
    + O(n^{-2/3})\right)\,,
    \label{eq:double-scaling-bn}
\end{equation}
with $s = c_0\, n^{2/3}\sigma$ and $d_1$ fixed by the local conformal data. This is the promised interpolation, and it degenerates correctly in both directions:
\begin{itemize}
    \item \textbf{$s\to-\infty$ (gap open, two-cut phase):} By \eqref{eq:HM-minus-infinity},
    $q(s)\sim\sqrt{-s/2} = n^{1/3}\sqrt{c_0|\sigma|/2}$, so the correction
    becomes $n$-independent,
    \begin{equation}
        b_n \;\longrightarrow\; b_\infty\Big(1 + (-1)^n d_1
        \sqrt{c_0|\sigma|/2}\Big)\,.
        \label{eq:PII-to-staggering}
    \end{equation}
    The Painlev\'e formula reproduces a finite even--odd staggering of
    amplitude proportional to $\sqrt{|\sigma|}\propto\epsilon_{\mathrm{gap}}$,
    which is precisely the $\Omega=\tfrac12$ two-cut behaviour
    \eqref{eq:ssh-stagger} with the amplitude \eqref{eq:ssh-exact-limits}
    proportional to the gap width.
    \item \textbf{$s\to+\infty$ (merged cut, single-cut phase):} Here 
    $q(s)\to\mathrm{Ai}(s)$ decays as
    $\tfrac{1}{2\sqrt\pi}s^{-1/4}e^{-\frac23 s^{3/2}}$, so the staggering
    switches off exponentially and the coefficients converge monotonically to
    $b_\infty$, as the single-cut result \eqref{eq:leading-asymptotics} for the
    merged support $[\omega_1^-,\omega_2^+]$ requires.
    \item \textbf{$s=0$ (critical point):} In this case, $q(0)\approx0.3679$ is a definite nonzero constant, and the staggering decays algebraically,
    \begin{equation}
        b_n = b_\infty\left(1 + \frac{(-1)^n d_1\, q(0)}{n^{1/3}}
        + O(n^{-2/3})\right).
        \label{eq:critical-bn}
    \end{equation}
\end{itemize}
To summarise; the Painlev\'e transcendent is the amplitude of the even--odd staggering of Section \ref{sec:multi-cut},  analytically continued through the point at which the gap closes.
Placing this beside the results of Sections \ref{sec:DZ} and \ref{sec:multi-cut} gives the following classification of subleading Lanczos behaviour by spectral topology:
\begin{center}
\begin{tabular}{lcc}
\hline
\textbf{Regime} & \textbf{Correction to $b_n$} & \textbf{Local parametrix} \\
\hline
Single cut, soft edges & $O(n^{-2})$ & Airy \\
Single cut, hard edges & $O(n^{-2})$ & Bessel \\
Two cuts, gap open & $O(1)$, oscillatory & Theta function \\
Gap closing & $O(n^{-1/3})$, staggered & Painlev\'e II \\
\hline
\end{tabular}
\end{center}
The critical correction is parametrically larger than either single-cut case, so convergence to the asymptotic Lanczos value is dramatically slower at the transition. The reason is essentially geometric; the density vanishes quadratically rather than as a square root, so the phase \eqref{eq:cubic-phase} is cubic rather than $3/2$-power, and the region over which the local model is needed widens from $n^{-2/3}$ to $n^{-1/3}$. A wider critical region admits a lower-order expansion, and the $\zeta^{-1}$ series \eqref{eq:PII-large-zeta} replaces the $\zeta^{-3/2}$ series of the Airy case.\\

\noindent
One caveat to this is that the analysis above assumes the merge is symmetric, so that $\Omega\to\tfrac12$ at the transition and the correction alternates with period two. This is the case treated in the matrix-model literature \cite{bleher2003double, claeys2006universality}, and the one for which the Painlev\'e II local model is established. For an asymmetric merge, where $\Omega_c\neq\tfrac12$, we expect the same $n^{-1/3}$ scale but a correction modulated at frequency $\Omega_c$ rather than staggered. We have not verified that the same local model applies, and we do not claim it here.

\subsection{Which gap closings are critical}
\label{subsec:which-transitions}

The SSH gap closes as $\Delta = |t_1-t_2|\to0$, and by \eqref{eq:ssh-exact-limits} the staggering amplitude $\tilde b = \Delta/2$ vanishes there, so the Lanczos sequence does indeed pass from staggered to monotone. The symmetry of the merge is likewise correct with $\omega_c = 0$  fixed by $\omega\to-\omega$ and $\Omega=\tfrac12$ throughout. However, as $\Delta\to0$ the numerator $|\omega|$ in \eqref{eq:ssh-measure} cancels the vanishing factor in $\sqrt{\omega^2-\Delta^2}$ exactly, leaving
\begin{equation}
    d\mu(\omega) \;\longrightarrow\;
    \frac{d\omega}{\pi\sqrt{\omega_{\max}^2-\omega^2}}\,,
    \label{eq:ssh-critical-limit}
\end{equation}
the arcsine measure, which is finite and \emph{nonzero} at $\omega_c=0$. Consequently the local condition \eqref{eq:quadratic-vanishing} does not hold. Far from acquiring a double zero, the density does not vanish there at all, and the merge point is an ordinary interior point of the merged support. To understand this, we turn to the quadratic map of Section \ref{subsec:solvable}. Under $x=\omega^2$ the SSH measure becomes the arcsine measure on $[\Delta^2,\omega_{\max}^2]$, which is single-cut for \emph{every} $\Delta$, including $\Delta=0$. In the $x$ variable no topology changes as the gap closes; an endpoint simply migrates to the origin, where it remains a hard edge governed by the Bessel parametrix of Section \ref{subsec:local-parametrices}. The genus-one curve in $\omega$ is a double cover of a genus-zero one, and the apparent gap closing in $\omega$ is the shadow of an endpoint collision with a symmetry-enforced fixed point in $x$. The transition is real and the staggering does disappear, but its local structure is Bessel, not Painlev\'e~II, and the relaxation is correspondingly $O(n^{-2})$ rather than $O(n^{-1/3})$.\\

\noindent
The distinction is a useful one, since both mechanisms remove a spectral gap and both terminate the staggering. What separates them is the behaviour of the density at the merge point. A double zero \eqref{eq:quadratic-vanishing} gives the Painlev\'e~II transition with its anomalous $n^{-1/3}$ relaxation, while a nonvanishing density gives an ordinary interior point and no critical slowing at all. Realising the former in a lattice model requires a family in which the two inner band edges meet with the density vanishing there, rather than one in which a symmetry-protected zero of the numerator conspires to cancel it. Constructing such a family within the chains of Section \ref{subsec:solvable}, and confirming the $n^{-1/3}$ scaling numerically, is a natural next step which we do not pursue here.

\subsection{Krylov criticality}
\label{subsec:krylov-criticality}

The exponent $1/3$ governs the rate at which the Krylov chain approaches translation invariance, and hence the duration of the transient before operator spreading reaches its asymptotic velocity. At the transition that approach is parametrically slower than in either adjacent phase, so the pre-asymptotic regime is correspondingly longer.
More broadly, a closing spectral gap is a qualitative change in the structure of the seed's Liouvillian spectrum, and in systems where the gap is controlled by a physical parameter like a coupling driving a transition from integrability to chaos, or a temperature opening or closing a gap in the density of states, the critical Lanczos behaviour is a sharp diagnostic of it. We call this a \emph{Krylov phase transition}; a change in the asymptotic structure of operator growth, visible in the Lanczos sequence as the decay of the staggering amplitude. At the critical point the relaxation is $n^{-1/3}$. The Hastings--McLeod solution of Painlev\'e~II interpolates between the two phases.\\

\noindent
The connection to matrix models is exact rather than an analogy. The one-cut to two-cut transition in the eigenvalue distribution of a random matrix ensemble is described by the same Painlev\'e~II double-scaling limit, with the same exponents and the same Hastings--McLeod solution \cite{bleher2003double, claeys2006universality}. From the Riemann--Hilbert side the mechanism is identical as branch points of the $g$-function collide, the genus of the spectral curve drops, and a Painlev\'e transcendent interpolates between the two topological phases. What the Krylov formulation adds is a dynamical interpretation. The same transition that reorganises an eigenvalue distribution reorganises the spreading of an operator in Krylov space, and the Painlev\'e transcendent governs the crossover between the two spreading behaviours.

\section{Refining the Operator Growth Hypothesis}
\label{sec:refined-universality}

Having developed the Riemann--Hilbert framework for
extracting asymptotic Lanczos data from spectral measures and applied it to four classes of measures of increasing complexity, in this
section we will use these results to revisit the operator growth hypothesis of~\cite{Parker:2018yvk}, reformulating it in the language of equilibrium
measures and orthogonal-polynomial universality. The main results are a refined classification of Krylov dynamics that goes beyond the leading
growth exponent, an explicit computation of subleading Lanczos
corrections for the chaotic spectral measures relevant to the SYK model, and a conjectural cross-class law for the prefactor correction, verified below in three exactly solvable families.\\

\subsection{Spectral tails and the leading growth law}
\label{subsec:spectral-tails}

The operator growth hypothesis of~\cite{Parker:2018yvk} proposes that chaotic quantum systems exhibit asymptotically linear Lanczos growth, $b_n \sim \alpha n$, with $\alpha$ playing a role analogous to a Lyapunov exponent. The argument connecting this growth law to the spectral measure proceeds via the observation that exponential tails in the spectral function, $\Phi(\omega) \sim e^{-|\omega|/\Lambda}$ for some scale $\Lambda$, produce linearly growing recurrence coefficients. The Riemann--Hilbert framework described above and in \cite{r9v1-nxj1} makes this connection precise and places it within a broader classification. Consider a spectral measure on $\mathbb{R}$ with density
\begin{equation}
\rho(\omega) \;\sim\; |\omega|^\gamma\,
e^{-c\,|\omega|^\beta}\,, \qquad |\omega| \to \infty\,,
\label{eq:freud-tail}
\end{equation}
where $\beta > 0$ controls the tail decay and $\gamma$ is a subleading power. For any such weight the orthogonal polynomials $\pi_n(\omega)$ concentrate their zeros on an interval whose length grows with $n$. The key step is the rescaling $\omega = n^{1/\beta}x$, which converts the fixed weight into varying form,
\begin{equation}
\rho(n^{1/\beta} x) \;\sim\;
n^{\gamma/\beta}\, |x|^\gamma\,
e^{-c\, n\, |x|^\beta}\,,
\label{eq:rescaled-weight}
\end{equation}
so that the essential $n$-dependence now sits in the exponent $e^{-nV(x)}$ with potential $V(x) = c|x|^\beta$. The algebraic factor $|x|^\gamma$ contributes only $(\gamma/n)\log|x|$ to the effective potential and therefore drops out of the equilibrium problem. It re-enters at first subleading order, and it is precisely there that the refinement of the operator growth hypothesis resides. The $x$-independent factor $n^{\gamma/\beta}$ is immaterial, since recurrence coefficients are invariant under overall rescaling of the weight.\\

\noindent
This is precisely the varying-weight setup of Section~\ref{subsec:gfunction}. Since $xV'(x) = c\beta|x|^\beta$ is increasing on $(0,\infty)$ for every $\beta > 0$, the equilibrium measure is supported on a single symmetric interval $[-a_\beta, a_\beta]$, and the endpoint condition~\eqref{eq:endpoint-conditions} evaluates in closed form to the Mhaskar--Rakhmanov--Saff number~\cite{MhaskarSaff1984} of the potential $c|x|^\beta$,
\begin{equation}
\frac{a_\beta}{2}
\;=\;
\left[\frac{\Gamma\!\left(\tfrac{\beta}{2}\right)
\Gamma\!\left(\tfrac{\beta}{2}+1\right)}
{c\;\Gamma(\beta+1)}\right]^{1/\beta}\,.
\label{eq:mrs-constant}
\end{equation}
The global parametrix then gives
\begin{equation}
b_n^{(\mathrm{rescaled})} \;\longrightarrow\;
\frac{a_\beta}{2}\,, \qquad n \to \infty\,,
\label{eq:rescaled-bn}
\end{equation}
for the rescaled polynomials. Undoing the rescaling, and restricting to $\beta \geq 1$ so that the underlying moment problem is determinate,\footnote{The algebraic identity~\eqref{eq:mrs-constant} holds for any $\beta > 0$, since the Gamma functions involved are finite there. The asymptotic statement~\eqref{eq:freud-growth} carries the additional hypothesis $\beta \geq 1$. For $\beta < 1$ the Krein condition
\begin{equation*}
\int \frac{\log(1/\rho(\omega))}{1+\omega^2}\,d\omega \;<\; \infty
\end{equation*}
is satisfied, the associated moment problem is indeterminate, and the correspondence between spectral measure and Lanczos sequence developed above is no longer one-to-one. See the discussion at the end of this subsection. Equation~\eqref{eq:mrs-constant} for $\beta < 1$ should therefore be read as a formal extremal-problem output, not as a physically meaningful growth rate.} the original recurrence coefficients satisfy
\begin{equation}
b_n \;\sim\;
\left[\frac{\Gamma\!\left(\tfrac{\beta}{2}\right)
\Gamma\!\left(\tfrac{\beta}{2}+1\right)}
{c\;\Gamma(\beta+1)}\right]^{1/\beta} n^{1/\beta}\,,
\qquad n \to \infty\,.
\label{eq:freud-growth}
\end{equation}
This is the \emph{Freud growth law}~\cite{Freud1976, LubinskyMhaskarSaff1988, kriecherbauer1999strong}. The leading asymptotic growth of the Lanczos coefficients is a pure power of $n$. Its exponent $1/\beta$ is determined solely by the tail decay rate of the spectral measure, and its coefficient by the equilibrium measure. Four cases deserve special attention:
\begin{itemize}
\item \textbf{Exponential tails} ($\beta = 1$):
$\rho(\omega) \sim e^{-|\omega|/\Lambda}$, \textit{i.e.}\ $c = 1/\Lambda$,
for which~\eqref{eq:mrs-constant} gives
$a_1/2 = \Gamma(\tfrac12)\Gamma(\tfrac32)\Lambda/\Gamma(2) = \pi\Lambda/2$, so that
\begin{equation}
b_n \;\sim\; \alpha\, n\,, \qquad \alpha = \frac{\pi\Lambda}{2}\,.
\label{eq:parker-slope}
\end{equation}
Equivalently $\Lambda = 2\alpha/\pi$, which is the slope--tail relation of~\cite{Parker:2018yvk}. The Riemann--Hilbert analysis therefore recovers the regime identified there as the universal signature of chaotic dynamics. The relation between spectral decay scale and Lanczos growth rate emerges as the $\beta=1$ member of the family~\eqref{eq:freud-growth}.
\item \textbf{Gaussian tails} ($\beta = 2$):
$\rho(\omega) \sim e^{-\omega^2/(2\sigma^2)}$, \textit{i.e.}\ $c = 1/(2\sigma^2)$, giving $a_2/2 = \sigma$ and $b_n \sim \sigma\sqrt{n}$. This is the Hermite class of Section~\ref{subsec:gaussian}.
\item \textbf{Quartic tails} ($\beta = 4$):
$\rho(\omega) \sim e^{-\omega^4}$ gives $(a_4/2)^4 = \Gamma(2)\Gamma(3)/\Gamma(5) = 1/12$, \textit{i.e.}\ $b_n \sim (n/12)^{1/4}$, reproducing the classical result of Nevai for the Freud weight $e^{-x^4}$~\cite{nevai1984asymptotics}. We record this case because it serves as a nontrivial check of the subleading analysis below.
\item \textbf{Compact support} ($\beta \to \infty$):
In the formal limit $\beta \to \infty$ the tail becomes a hard cutoff and $n^{1/\beta} \to 1$, recovering the constant asymptotics $b_n \to (B-A)/4$ of Section~\ref{subsec:global-parametrix}.
\end{itemize}

\noindent
Within the family~\eqref{eq:freud-tail} of stretched-exponential tails, the leading growth exponent therefore defines a one-parameter family of universality classes indexed by $\beta$. Three qualifications delimit the scope of this classification.\\

\noindent
First, it applies to tails of pure stretched-exponential form. Slowly varying modifications, $\rho \sim e^{-c|\omega|^\beta \ell(|\omega|)}$ with $\ell$ slowly varying, produce corresponding slowly varying corrections to the growth law. The physically important example is a one-dimensional local system, for which $b_n \sim \alpha n/\ln n$~\cite{Parker:2018yvk}. The logarithm places it outside the family~\eqref{eq:freud-tail}. It should be regarded as a marginal deformation of the $\beta = 1$ class rather than a separate class, and every statement in this section is subject to the same qualification.\\

\noindent
Second, for $\beta < 1$ the Krein condition $\int \log(1/\rho(\omega))(1+\omega^2)^{-1}d\omega < \infty$ is satisfied and the moment problem for $\rho$ is indeterminate. The Jacobi matrix assembled from $\{a_n, b_n\}$ is then in the limit-circle case at infinity, and distinct measures share identical Lanczos data. The Liouvillian of a genuine quantum system is of course self-adjoint and selects a unique measure, but that measure can no longer be reconstructed from the Lanczos coefficients alone. Relatedly, the super-linear growth $b_n \sim n^{1/\beta}$ with $1/\beta > 1$ transports the semiclassical Krylov wavefront, $\dot{n} = 2b_n$, to $n = \infty$ in finite time. The true unitary dynamics is then sensitive to the self-adjoint extension, that is to a boundary condition at the far end of the Krylov chain. We therefore regard the $\beta < 1$ classes as formal.\\

\noindent
Third, the classification concerns the leading power alone. Measures sharing $\beta$ are distinguished by the finer invariants to which we now turn.

\subsection{Subleading corrections and the Meixner--Pollaczek family}
\label{subsec:subleading-chaotic}

The leading growth law $b_n \sim \alpha n$ is indeed universal for all exponential-tail measures. The Riemann--Hilbert framework reveals that the subleading Lanczos structure carries additional physical information invisible to the tail-based argument. To make this concrete, in this section we compute the full subleading expansion for a family that includes the conformal limit of the SYK model~\cite{Maldacena:2016hyu}. toward this end, consider the spectral densities
\begin{equation}
w_\lambda(\omega) \;=\; \frac{1}{2\pi}\,
\left|\Gamma\!\left(\lambda + \frac{i\omega}{\Lambda}
\right)\right|^2, \qquad \lambda > 0\,,
\label{eq:MP-weight}
\end{equation}
on $\mathbb{R}$, with $\Lambda > 0$ setting the energy scale. By Stirling's formula, the large-$|\omega|$ behaviour is
\begin{equation}
w_\lambda(\omega) \;\sim\;
\left|\frac{\omega}{\Lambda}\right|^{2\lambda - 1}\,
e^{-\pi|\omega|/\Lambda}\,, \qquad |\omega| \to \infty\,,
\label{eq:MP-tail}
\end{equation}
so this is an exponential-tail weight with $\beta = 1$ and a power-law prefactor controlled by $\lambda$. In the conventions of Section~\ref{subsec:spectral-tails} the decay scale is $\Lambda/\pi$, so the Freud law~\eqref{eq:parker-slope} predicts $\alpha = \tfrac{\pi}{2}\cdot\tfrac{\Lambda}{\pi} = \Lambda/2$. The exact recurrence below confirms this, providing an independent check of~\eqref{eq:mrs-constant}. For $\lambda = 1/2$ the reflection formula gives $|\Gamma(1/2 + ix)|^2 = \pi/\cosh(\pi x)$, so that $w_{1/2}(\omega) = \tfrac12\,\mathrm{sech}(\pi\omega/\Lambda)$, a pure $\mathrm{sech}$ weight with no power-law prefactor.\\

\noindent
The orthogonal polynomials with respect to $w_\lambda$ are the \emph{Meixner--Pollaczek} polynomials $P_n^{(\lambda)}(\omega/\Lambda; \pi/2)$~\cite{koekoek2010hypergeometric}, for which the monic three-term recurrence $\omega\, \pi_n = \pi_{n+1} + a_n \pi_n + \beta_n \pi_{n-1}$ has coefficients
\begin{equation}
a_n \;=\; 0\,, \qquad
\beta_n \;=\; \frac{\Lambda^2}{4}\, n(n + 2\lambda - 1)\,.
\label{eq:MP-recurrence}
\end{equation}
The diagonal coefficients vanish by the $\omega \to -\omega$ symmetry of the weight. The off-diagonal Lanczos coefficients are therefore
\begin{equation}
b_n \;=\; \sqrt{\beta_n} \;=\;
\frac{\Lambda}{2}\,\sqrt{n(n + 2\lambda - 1)}\,,
\label{eq:MP-bn-exact}
\end{equation}
and exact for all $n \geq 1$. Expanding for large $n$ as,
\begin{equation}
b_n \;=\; \frac{\Lambda}{2}\, n\,
\sqrt{1 + \frac{2\lambda - 1}{n}}
\;=\; \frac{\Lambda}{2}\, n
\left(1 + \frac{2\lambda - 1}{2n}
- \frac{(2\lambda - 1)^2}{8n^2}
+ O(n^{-3})\right),
\label{eq:MP-expansion}
\end{equation}
gives the asymptotic series
\begin{equation}
b_n \;=\; \alpha\, n \;+\; b_0
\;+\; \frac{b_1}{n} \;+\; O(n^{-2})\,,
\label{eq:chaotic-subleading}
\end{equation}
with
\begin{equation}
\alpha \;=\; \frac{\Lambda}{2}\,, \quad
b_0 \;=\; \frac{\Lambda(2\lambda - 1)}{4}
\;=\; \alpha\!\left(\lambda - \tfrac{1}{2}\right),
\quad
b_1 \;=\; -\frac{\Lambda(2\lambda - 1)^2}{16}
\;=\; -\frac{\alpha(2\lambda - 1)^2}{8}\,.
\label{eq:subleading-coefficients}
\end{equation}
Four features of this expansion are worth noting.
\begin{enumerate}
\item The leading rate $\alpha = \Lambda/2$ depends only on the exponential decay scale and is insensitive to $\lambda$, consistent with the tail-based argument of~\cite{Parker:2018yvk}. All weights in the family $\{w_\lambda\}_{\lambda > 0}$ share the same leading growth.
\item The constant offset $b_0 = \alpha(\lambda - 1/2)$ is the first subleading invariant. It depends on the power-law prefactor through $\lambda$ and vanishes at $\lambda = 1/2$. Two systems with the same decay scale but different $\lambda$ share identical leading growth and are distinguished by their constant offsets. This is a concrete example of information beyond the reach of the spectral-tail argument.
\item The sign of $b_0$ is fixed by $\lambda$, positive for $\lambda > 1/2$ and negative for $\lambda < 1/2$. The Lanczos sequence therefore approaches its asymptotic growth from above or below according to the power-law structure of the tails. At $\lambda = 1/2$, where $b_n = \alpha n$ exactly with no subleading corrections at all, the weight is the pure $\mathrm{sech}$.
\item The higher coefficients $b_1, b_2, \ldots$ carry further information but are suppressed by powers of $n$. The full tower $\{b_0, b_1, b_2, \ldots\}$ is determined by the single parameter $\lambda$.
\end{enumerate}

\noindent
From the Riemann--Hilbert perspective the structure of~\eqref{eq:chaotic-subleading} is organised by the effective potential. Under the rescaling $\omega = nx$, Stirling's formula gives
\begin{equation}
V_{\mathrm{eff}}(x) \;=\; -\frac{1}{n}\log w_\lambda(nx)
\;=\; \frac{\pi|x|}{\Lambda}
\;-\; \frac{2\lambda - 1}{n}\,\log\frac{n|x|}{\Lambda}
\;+\; O(n^{-1})\,,
\label{eq:effective-potential}
\end{equation}
up to $x$-independent terms, which do not affect the recurrence coefficients. Two effects enter at order $1/n$, and both are localised at the origin. This is an \emph{interior} point of the support $[-a_1, a_1]$, at which the equilibrium density of the linear potential is strictly positive. They are the logarithmic perturbation $-\tfrac{2\lambda-1}{n}\log|x|$ of the potential, and the smoothing of the cusp of $\pi|x|/\Lambda$ on the scale $|x| \lesssim 1/n$, where the Stirling form fails and the exact weight is analytic. Neither effect displaces the endpoints at leading order, which is why $\alpha$ is insensitive to $\lambda$. Both feed through the small-norm expansion as an $O(1/n)$ relative correction to the rescaled recurrence coefficient, with coefficient $\lambda - 1/2$. At $\lambda = 1/2$ the logarithmic perturbation is absent and the smoothed cusp of the $\mathrm{sech}$ weight leaves no residue, consistent with the exactness of $b_n = \alpha n$ there.\\

\noindent
We should be clear about the status of these statements. Equation~\eqref{eq:MP-bn-exact} and its expansion are exact consequences of the classical Meixner--Pollaczek recurrence. The Riemann--Hilbert mechanism just described explains why $b_0$ should depend only on the tail data $(\Lambda, \lambda)$ and not on the interior shape of the weight. However, the local parametrix at the smoothed cusp, which is what a proof of within-class universality of $b_0$ would require, is not constructed in this paper. Universality of the offset beyond the Meixner--Pollaczek family is therefore a conjecture, albeit one with independent support across tail classes, as we now describe.

\subsubsection*{A cross-class law for the prefactor correction}

The Meixner--Pollaczek computation determines the $1/n$ relative correction in the $\beta = 1$ class. Exactly solvable families at other values of $\beta$ allow the pattern to be tested across classes. Throughout, we write $1\!\!1[n\ \mathrm{odd}]$ for the indicator function equal to $1$ when $n$ is odd and $0$ when $n$ is even.\\

\noindent
For the generalised Hermite weight $w(\omega) = |\omega|^{2\mu}e^{-\omega^2}$ ($\beta = 2$, $\gamma = 2\mu$), the recurrence coefficients are classical~\cite{chihara2011introduction}. In the convention $b_n^2 = \beta_n$,
\begin{equation}
\beta_{2m} = \frac{m}{2}\,,
\qquad
\beta_{2m+1} = m + \mu + \tfrac12\,,
\qquad\text{that is}\qquad
b_n^2 = \frac{n}{2} + \mu\,1\!\!1[n\ \mathrm{odd}]\,,
\label{eq:genhermite-beta}
\end{equation}
so that
\begin{equation}
b_n \;=\; \sqrt{\tfrac{n}{2}}
\left(1 + \frac{\mu}{2n} - \frac{\mu}{2n}(-1)^n
+ O(n^{-2})\right).
\label{eq:genhermite-bn}
\end{equation}
The smooth and alternating amplitudes are equal and opposite here, so the $1/n$ correction cancels identically on even $n$ and doubles on odd $n$. This is the $\beta=2$ instance of a pattern we return to below.
For the quartic Freud weight $w(\omega) = |\omega|^{\rho}e^{-\omega^4}$ ($\beta = 4$, $\gamma = \rho$), the string equation~\cite{nevai1984asymptotics}
\begin{equation}
4\,\beta_n\left(\beta_{n-1} + \beta_n + \beta_{n+1}\right)
\;=\; n + \rho\,1\!\!1[n\ \mathrm{odd}]
\label{eq:quartic-string}
\end{equation}
may be solved asymptotically with the ansatz $\beta_n = \sqrt{n/12}\,(1 + u_n)$, $u_n = (s + p(-1)^n)/n$. Matching the smooth and alternating components separately gives $s = \rho/4$ and $p = -3\rho/4$. The neighbour coupling in~\eqref{eq:quartic-string} triples the oscillation amplitude relative to the naive substitution $n \to n + \rho\,1\!\!1[n\ \mathrm{odd}]$, and hence
\begin{equation}
b_n \;=\; \left(\tfrac{n}{12}\right)^{1/4}
\left(1 + \frac{\rho}{8n} - \frac{3\rho}{8n}(-1)^n
+ O(n^{-2})\right).
\label{eq:quartic-bn}
\end{equation}
The leading constant $(1/12)^{1/4}$ agrees with the Mhaskar--Rakhmanov--Saff formula~\eqref{eq:mrs-constant}, providing a consistency check between the string equation and the equilibrium problem.\\

\noindent
The parity-averaged relative coefficient of $1/n$ in these three families is $\gamma/2$ ($\beta = 1$), $\gamma/4$ ($\beta = 2$) and $\gamma/8$ ($\beta = 4$). All three obey a single law, which we conjecture to hold generally; for a spectral density $\rho(\omega) \sim |\omega|^\gamma e^{-(|\omega|/\omega_0)^\beta}$,
\begin{equation}
b_n \;=\;
\left[\frac{\Gamma(\tfrac\beta2)\Gamma(\tfrac\beta2+1)}
{\Gamma(\beta+1)}\right]^{1/\beta}
\omega_0\, n^{1/\beta}
\left(1 \;+\; \frac{\gamma}{2\beta\, n}
\;+\; \frac{A_{\mathrm{osc}}}{n}\,(-1)^n
\;+\; o(n^{-1})\right).
\label{eq:prefactor-law}
\end{equation}
At $\beta = 1$ the smooth term reproduces $b_0 = \alpha\gamma/2 = \alpha(\lambda - \tfrac12)$.
The oscillatory amplitude $A_{\mathrm{osc}}$ vanishes for the Meixner--Pollaczek family, whose weight is smooth and positive on all of $\mathbb{R}$. It is nonzero in the two families above, whose algebraic factor $|\omega|^\gamma$ is a genuine singularity of the measure at an interior point of the rescaled support. The natural interpretation, in the language of Section~\ref{sec:multi-cut}, is that an interior algebraic singularity acts as a spectral gap of zero width. The induced modulation has frequency $2\pi\,\mu_{\mathrm{eq}}([-a_\beta, 0]) = \pi$ by symmetry, \textit{i.e.}\ $\cos(\pi n) = (-1)^n$, in exact parallel with the filling-fraction formula~\eqref{eq:oscillatory-bn}, but with amplitude decaying as $1/n$ rather than $O(1)$. This is commensurate with the vanishing width.\\

\noindent
We further record, without claiming it, that the measured amplitudes $A_{\mathrm{osc}} = -\gamma/4$ ($\beta = 2$) and $-3\gamma/8$ ($\beta = 4$) are both consistent with
\begin{equation}
A_{\mathrm{osc}} \;=\; -\frac{(\beta - 1)\,\gamma}{2\beta}\,,
\label{eq:osc-conjecture}
\end{equation}
which would in addition explain the absence of oscillation at $\beta = 1$ independently of the smoothness argument. Distinguishing the two explanations requires a genuinely singular $\beta = 1$ weight, which we have not been able to treat exactly. We leave it as an open problem, together with the Riemann--Hilbert proof of~\eqref{eq:prefactor-law} via the interior local parametrix.

\subsubsection*{Application: the SYK model}

In the Sachdev--Ye--Kitaev model~\cite{Sachdev1993,Kitaev2015} the spectral function of the fundamental fermion in the conformal limit, computed with respect to the Wightman (thermally symmetrised) inner product used in~\cite{Parker:2018yvk}, is the Fourier transform of $G^W(t) \propto \left[\pi/(\beta\cosh(\pi t/\beta))\right]^{2\Delta}$. It takes the form~\cite{Maldacena:2016hyu}
\begin{equation}
\Phi_{\mathrm{SYK}}(\omega) \;\propto\;
\left|\Gamma\!\left(\Delta + \frac{i\beta\omega}{2\pi}
\right)\right|^2,
\label{eq:SYK-spectral}
\end{equation}
where $\Delta = 1/q$ is the scaling dimension in $\mathrm{SYK}_q$ and $\beta$ is the inverse temperature. This power spectrum is even in $\omega$, so $a_n = 0$. It is precisely the Meixner--Pollaczek weight~\eqref{eq:MP-weight} with $\lambda = \Delta = 1/q$ and $\Lambda = 2\pi/\beta = 2\pi T$. The Lanczos coefficients are therefore given by~\eqref{eq:MP-bn-exact},
\begin{equation}
b_n^{(\mathrm{SYK}_q)} \;=\; \pi T\,
\sqrt{n\!\left(n + \frac{2}{q} - 1\right)}\,.
\label{eq:SYK-bn}
\end{equation}
The leading growth rate $\alpha = \pi T$ is independent of $q$ and depends only on the temperature, consistent with the operator growth hypothesis. Since the SYK model is maximally chaotic, $\lambda_L = 2\pi T$, saturating the Maldacena-Shenker-Stanford bound $\lambda_L \leq 2\alpha$ of~\cite{Maldacena:2015waa}. In the present context, the Riemann--Hilbert computation recovers maximal-chaos saturation directly from the $\Gamma$-function tail.\\

\noindent
The subleading constant
\begin{equation}
b_0^{(\mathrm{SYK}_q)} \;=\;
\pi T\!\left(\frac{1}{q} - \frac{1}{2}\right)
\label{eq:SYK-b0}
\end{equation}
depends on $q$ and is negative for $q > 2$. Three cases are worth recording.
\begin{itemize}
\item $q = 2$: the formal value $\Delta = 1/2$ gives $b_0 = 0$ and $b_n = \pi T n$ exactly, the $\mathrm{sech}$ point of the family. This point should \emph{not} be identified with physical $\mathrm{SYK}_2$. The free model has no conformal regime, its Liouvillian frequencies are single-particle energies drawn from a compactly supported semicircular distribution, and its true Lanczos coefficients are bounded, placing it in the single-cut class of Section~\ref{subsec:global-parametrix}. The $\Delta \to 1/2$ member of the conformal family and the free theory are distinct spectral measures that happen to carry the same value of $q$.
\item $q = 4$: $\Delta = 1/4$ and $b_0 = -\pi T/4$. The negative offset means the Lanczos coefficients approach their asymptotic slope from below.
\item $q \to \infty$: $\Delta \to 0$, $b_0 \to -\pi T/2$, and $b_n \to \pi T\sqrt{n(n-1)} = \pi T(n - 1/2 + O(n^{-1}))$. This limit is formal. At $\Delta = 0$ the weight loses integrability at the origin, and $b_1 = \pi T\sqrt{2/q} \to 0$, signalling the decoupling of the seed from the rest of the Krylov chain. The physical large-$q$ analysis retains the $1/q$ corrections.
\end{itemize}
The scaling dimension $\Delta$ is therefore a subleading Lanczos invariant. It leaves no trace in the leading growth rate but is extracted from the constant offset $b_0$. In Krylov terms, operators of different scaling dimensions in the SYK model spread at the same asymptotic rate but differ in their transient behaviour, with $b_0$ controlling the duration and character of the pre-asymptotic regime.

\subsection{A refined universality classification}
\label{subsec:refined-classification}

The computations above illustrate a general principle. The leading growth law~\eqref{eq:freud-growth} is the coarsest invariant of the asymptotic Lanczos structure, the analogue of a Lyapunov exponent. The Riemann--Hilbert analysis reveals a hierarchy of finer invariants, which we organise into four levels ordered by the size of the effect each produces.

\begin{enumerate}
\item \textbf{Tail exponent:}
The exponent $\beta$ in the spectral tail $\rho \sim e^{-c|\omega|^\beta}$ determines the leading Lanczos growth $b_n \sim n^{1/\beta}$, with the explicit coefficient~\eqref{eq:mrs-constant}. This is the content of the Freud growth law, and it subsumes the classification of~\cite{Parker:2018yvk} as the $\beta = 1$ case, within the scope delimited in Section~\ref{subsec:spectral-tails}.

\item \textbf{Spectral topology:}
The number of connected components of the support, and the filling fractions of its bands, determine whether the Lanczos sequence converges or oscillates. A measure with $p$ gaps produces quasiperiodic modulation at $p$ independent frequencies $\Omega_j$, fixed by the support alone through~\eqref{eq:filling-fraction}. For compactly supported measures this modulation enters at order one, the same order as the leading constant itself. It is therefore beyond-tail data rather than subleading data in the strict sense, and it is the coarsest invariant not visible to Level~1.

\item \textbf{Local spectral data:}
Within a fixed tail class and a fixed topology, the Riemann--Hilbert analysis resolves further structure. The prefactor exponent $\gamma$ controls the universal relative correction $\gamma/(2\beta n)$ of the conjectured behaviour~\eqref{eq:prefactor-law}. This is realised in the chaotic class as the constant offset $b_0 = \alpha\gamma/2$, together with a parity-alternating component whose presence registers whether the prefactor is a genuine interior singularity of the measure or just its tail asymptotics. Independently, the endpoint exponents distinguish soft from hard edges and control the correction series at order $n^{-2}$ (as described in Section~\ref{subsec:local-parametrices}). Two systems sharing a leading growth rate and a topology but differing at this level are asymptotically equivalent at leading order yet exhibit measurably different Lanczos sequences. In the SYK context this is the statement that operators of different scaling dimension share the Lyapunov-like exponent $\alpha$ but are separated by their subleading Krylov data.

\item \textbf{Critical structure.}
Near a spectral phase transition the subleading corrections are enhanced and governed by Painlev\'e transcendents (Section~\ref{sec:critical-krylov}). Systems approaching the same type of transition share the same critical exponent. Specifically, this means an $n^{-1/3}$ scaling for the gap-closing class, where the equilibrium density acquires a double zero at the merge point and the crossover is controlled by the Hastings--McLeod solution of Painlev\'e~II. Transitions with a different local structure at the merge point scale differently, and Section~\ref{subsec:which-transitions} exhibits one such case, in which a symmetry-protected merge leaves the density nonvanishing and the relaxation reverts to the Bessel value $n^{-2}$. This level is the finest of the four and is relevant whenever a physical parameter tunes the spectral measure through a topological transition.
\end{enumerate}

\noindent
The four levels are ordered by the magnitude of the effect; $O(n^{1/\beta})$ for the growth law, $O(n^{1/\beta})$ modulation for topology, $O(n^{1/\beta - 1})$ for the local data, and enhanced algebraic relaxation at criticality. The operator growth hypothesis of~\cite{Parker:2018yvk} operates at Level~1 alone. Everything below it is invisible to spectral tails, and accessible through the global geometry of the spectral measure.

\section{Discussion}
\label{sec:discussion}

The asymptotics of Krylov dynamics can be mapped to the asymptotics of a Riemann--Hilbert
problem \cite{r9v1-nxj1}. Once the spectral measure of the seed operator is identified with the
orthogonality weight of a Fokas--Its--Kitaev problem, the Lanczos coefficients sit in the large-$z$ expansion of its solution, and the Deift--Zhou steepest-descent method converts questions about operator growth into questions
about equilibrium measures, parametrices, and small-norm expansions. The main
claim of this article is what that conversion then reveals; the global topology
of the spectral measure, \textit{i.e.} how many components its support has, how they are
separated, and what happens when they merge, controls the Lanczos sequence at
and beyond leading order, and does so in ways no property of the spectral tail
can see. The output is a hierarchy of asymptotic Lanczos data considerably finer
than the leading growth exponent. Ordered by the size of the effect:

\begin{itemize}
\item At leading order, the coefficients are fixed by the support of the
equilibrium measure alone. For compactly supported measures this is the band
edges while, for stretched-exponential tails it is the Mhaskar--Rakhmanov--Saff
number \eqref{eq:mrs-constant}. The growth law of the operator growth hypothesis \cite{Parker:2018yvk} is recovered as the $\beta=1$ member of the
Freud family \eqref{eq:freud-growth}.
\item The topology of the support enters at the same order as the leading
constant itself. A multi-cut measure produces the quasiperiodic modulation
\eqref{eq:oscillatory-bn}, with frequency fixed by the filling fraction alone,
and a gap-closing transition produces the Painlev\'e~II crossover
\eqref{eq:double-scaling-bn} with its anomalously slow $n^{-1/3}$ relaxation.
The latter is slower than the relaxation of either adjacent phase.
\item At subleading order, the coefficients encode local data of the
measure. These include the endpoint densities at soft edges, the endpoint
exponents at hard edges, and, in the chaotic class, the power-law prefactor of
the spectral tail. The prefactor appears as the constant offset
$b_0 = \alpha\gamma/2$, and for the conformal SYK measure it encodes the
operator dimension through $b_0 = \pi T(\Delta - \tfrac12)$.
\end{itemize}

\noindent
We exhibited each level of this hierarchy in a solvable setting. The SSH chain realizes the two-cut theta-function structure at the rational point
$\Omega=\tfrac12$. Its next-nearest-neighbour deformation realizes the generic quasiperiodic regime, with the oscillation frequency predicted by the harmonic measure of the support. The Meixner--Pollaczek family realizes the subleading structure of the chaotic class in closed form. The
practical moral is that two systems agreeing on $\alpha$, or more generally
on the tail exponent $\beta$, need not be asymptotically equivalent as Krylov problems, and the data distinguishing them is computable.\\

\noindent
It is worth being precise about the status of the various statements made here. The single-cut analysis of Section~\ref{sec:DZ} and Appendix~\ref{app:steepest-descent} is a theorem of the orthogonal-polynomial literature; our contribution here is its transcription into Krylov language. The multi-cut and critical analyses of Sections~\ref{sec:multi-cut} and~\ref{sec:critical-krylov} likewise rest on classical machinery, the theta-function parametrix and Painlev\'e~II double-scaling limit of \cite{bleher2003double, claeys2006universality}. What is new is the application. Specifically, that the oscillation frequency of a Lanczos sequence is a support-only invariant, and hence predictable in advance of any computation of the dynamics, appears not to have been noted. Nor has the gap-closing transition previously been read as a transition in operator growth.\\

\noindent
Two statements, both in Section~\ref{sec:refined-universality}, remain conjectural. The first is the cross-class law \eqref{eq:prefactor-law} for the $1/n$ correction, which we have verified in three exactly solvable families ($\beta = 1, 2, 4$) but not proven. The second is the candidate amplitude $A_{\rm osc} = -(\beta-1)\gamma/(2\beta)$ for the parity-alternating component. A proof of either requires the construction of a local parametrix at an interior algebraic singularity of the weight, a confluent hypergeometric model problem which we have not carried out. A genuinely singular $\beta=1$ weight would in addition separate the two competing explanations for the vanishing of $A_{\rm osc}$ in the Meixner--Pollaczek family. Our statements about the Krylov complexity $K(t)$ itself are also weaker than those about the coefficients. The ballistic growth with beating envelope in the multi-cut case, and the prolonged transient at criticality, are semiclassical inferences from the coefficient asymptotics rather than controlled asymptotics of the wavefunction $\phi_n(t)$.\\

\noindent
Several directions seem worth pursuing further:

\begin{itemize}
    \item The Deift--Zhou method was invented for oscillatory Riemann--Hilbert problems in \emph{time-dependent} settings, and there is a natural formulation in which the Krylov wavefunction
$\phi_n(t)$, rather than the Lanczos coefficients, is the primary object. The integral representation~\eqref{eq:int-rep} is amenable to a joint large-$(n,t)$ steepest-descent analysis with $n/t$ held fixed. This would upgrade the semiclassical statements (see also \cite{Murugan:2026yyu, Bhattacharyya:2026qef}) about $K(t)$ to
controlled asymptotics with error terms, and would resolve the wavefront structure that the coefficient asymptotics alone cannot see.
\item  Open-system generalizations of
Krylov complexity replace the Liouvillian by a Lindbladian, and the Lanczos recursion by its bi-orthogonal variant. The natural analytic home
for bi-orthogonal polynomials with complex weights is again a
Riemann--Hilbert problem, but one without positivity, for which existence of the solution 
is no longer automatic. The breakdown points of the bi-orthogonal recursion may themselves carry physical meaning for dissipative operator growth. The steepest-descent analysis of such problems is an open direction with substantial precedent in the non-Hermitian random matrix literature.
\item We analyzed the gap-closing
transition, governed by Painlev\'e~II. The complementary transition, the
birth of a new cut, in which an isolated band nucleates at a local minimum of
the effective potential, is resolved by a different local analysis and produces different critical scaling. Section~\ref{subsec:which-transitions} exhibits a third possibility, in which a symmetry-protected merge leaves the density nonvanishing at the merge point and the relaxation reverts to the Bessel value. A complete taxonomy of Krylov phase
transitions would enumerate the topology-changing processes of the
equilibrium measure together with their transcendents and exponents, in parallel with the classification familiar from unitary matrix models.
\item In random settings the natural object is the ensemble-averaged Lanczos sequence. The results of \cite{Qu:2025lgo} suggest that the present single-measure analysis admits an averaged counterpart in the large-$N$ limit, and fluctuation statistics
of the $b_n$ around the Riemann--Hilbert prediction would then furnish a finer probe of spectral rigidity than the mean growth rate.
\item Since the
subleading invariant $b_0$ resolves the operator dimension in the conformal SYK measure, it is natural to ask what bulk quantity it computes in a
holographic dual, and whether the hierarchy of Lanczos data assembled here has
a geometric counterpart organized, like the leading term, around the horizon. A companion paper \cite{Graef:2026pzv} addresses the boundary side of this question in $\mathcal{N}=4$ super Yang--Mills, where finite-density probe states fall in the Hermite class of Section~\ref{subsec:gaussian}. The translation of that analysis from state to operator Krylov complexity, and hence into the framework developed here, is the subject of work in progress.
\end{itemize}

\noindent
We began this work by noting the particular satisfaction of finding an old corner of mathematics waiting for a new physics problem. We end by observing that this particular corner is deeper than it first appears. The same Riemann--Hilbert machinery that fixes the leading growth of operators also resolves their oscillations, their subleading structure, and the transitions between them.

\acknowledgments
We would like to thank Eric Graef and Horatiu Nastase for collaboration on related ideas, and Pawel Caputa whose talk at YITP was the inspiration for the idea behind this work. JM would like to acknowledge Kathy Driver for many interesting discussions over the years and regrets that it took him so long to write a paper on orthogonal polynomials. We would like to thank the organisers of the Holographic Universe 2026 workshop at the YITP, Kyoto where this work was conceived. JM would also like to thank the IFT and ICTP-SAIFR for their warm hospitality during which much of this work was completed. JM and HJRVZ are supported in part by the ``Quantum Technologies for Sustainable Development" grant
from the National Institute for Theoretical and Computational Sciences of South Africa
(NITheCS). 
MW is supported by a Grant-in-Aid for JSPS Fellows No.~22KJ1777 and a Grant-in-Aid for Early-Career Scientists No.~25K17387.

\appendices
\section{Verification of the Fokas--Its--Kitaev problem}
\label{app:RH-verification}

We verify here that \eqref{eq:Y-def} solves the Riemann--Hilbert problem stated in Section \ref{sec:RH-OP}, and that the solution is unique.

\subsection*{Existence}

The first column of $Y_n$ consists of polynomials and is therefore entire; the second column consists of Cauchy transforms and is analytic off $\mathbb{R}$. This establishes condition 1. For the jump condition, the Sokhotski--Plemelj relation gives
\begin{equation}
    \mathcal{C}_+[\pi_n w](x) = \mathcal{C}_-[\pi_n w](x) + \pi_n(x)w(x)\,,
\end{equation}
and identically for $\pi_{n-1}$. Since the first column has no jump, this is precisely
\begin{equation}
    Y_{n,+}(x) = Y_{n,-}(x)
    \begin{pmatrix} 1 & w(x) \\ 0 & 1 \end{pmatrix}\,,
\end{equation}
which is \eqref{eq:Y-jump}.
The normalisation at infinity follows from monicity and orthogonality. Expanding the Cauchy transform in powers of $1/z$ gives
\begin{equation}
    \mathcal{C}[\pi_n w](z) = -\frac{1}{2\pi i}
    \sum_{k=0}^{\infty}\frac{1}{z^{k+1}}
    \int_{\mathbb{R}} s^{k}\pi_n(s)w(s)\, ds\,,
    \label{eq:cauchy-expansion}
\end{equation}
and using orthogonality of $\pi_n$ against every polynomial of degree less than $n$,
\begin{equation}
    \int_{\mathbb{R}} s^{k}\pi_n(s)w(s)\, ds = 0\,,
    \qquad k = 0,\ldots,n-1\,,
\end{equation}
the first nonvanishing term in \eqref{eq:cauchy-expansion} occurs at $k = n$. Since $\pi_n$ is monic, $s^n = \pi_n(s) + (\text{lower degree})$, so
\begin{equation}
    \int_{\mathbb{R}} s^{n}\pi_n(s)w(s)\, ds = h_n\,,
\end{equation}
and therefore
\begin{equation}
    \mathcal{C}[\pi_n w](z) = -\frac{h_n}{2\pi i}z^{-n-1}
    + O(z^{-n-2})\,,
    \qquad
    \mathcal{C}[\pi_{n-1}w](z) = -\frac{h_{n-1}}{2\pi i}z^{-n}
    + O(z^{-n-1})\,,
\end{equation}
which is \eqref{eq:cauchy-leading}. Assembling the four entries gives
$Y_n(z) = (\mathbb{I} + O(z^{-1}))\, z^{n\sigma_3}$, establishing condition 3.

\subsection*{Uniqueness}

Conversely, suppose $Y_n$ solves the problem. The jump matrix in \eqref{eq:Y-jump} is upper triangular with unit diagonal, so the first column has no jump across $\mathbb{R}$ and $Y_{11}, Y_{21}$ extend to entire functions. The normalisation \eqref{eq:Y-normalization} then forces
\begin{equation}
    Y_{11}(z) = z^{n} + O(z^{n-1})\,,
    \qquad
    Y_{21}(z) = O(z^{n-1})\,,
\end{equation}
so $Y_{11}$ is a monic polynomial of degree $n$ and $Y_{21}$ a polynomial of degree at most $n-1$. The jump condition on the upper-right entry reads
\begin{equation}
    Y_{12,+}(x) - Y_{12,-}(x) = Y_{11}(x)w(x)\,,
\end{equation}
so $Y_{12}$ differs from $\mathcal{C}[Y_{11}w]$ by an entire function, which the normalisation at infinity forces to vanish; hence $Y_{12} = \mathcal{C}[Y_{11}w]$. The asymptotic condition $Y_{12}(z) = O(z^{-n-1})$ combined with the expansion \eqref{eq:cauchy-expansion} then requires
\begin{equation}
    \int_{\mathbb{R}} x^{k}Y_{11}(x)w(x)\, dx = 0\,,
    \qquad k = 0,\ldots,n-1\,,
\end{equation}
so $Y_{11}$ is the monic orthogonal polynomial $\pi_n$. An identical argument on the second row identifies $Y_{21}$ with $-2\pi i\, h_{n-1}^{-1}\pi_{n-1}$, establishing the equivalence of the orthogonal-polynomial and Riemann--Hilbert problems.

\section{Steepest descent for single-cut measures}
\label{app:steepest-descent}

This appendix collects the constructions summarised in Section \ref{sec:DZ}, specifically, the resolvent computation of the equilibrium measure, the Airy and Bessel local parametrices, and the small-norm analysis that produces the correction series. We follow \cite{deift1999orthogonal} throughout; but see also \cite{r9v1-nxj1} for a treatment adapted to the Krylov setting, including a careful account of the regularity hypotheses on the weight.

\subsection{The equilibrium measure from the resolvent}
\label{app:equilibrium}

Introduce the resolvent of the equilibrium measure,
\begin{equation}
    \omega(z) \equiv \int_A^B dy\,
    \frac{\rho_{\mathrm{eq}}(y)}{z - y}\,,
    \qquad z \in \mathbb{C}\setminus[A,B]\,.
    \label{eq:resolvent-def}
\end{equation}
The Euler--Lagrange equation \eqref{eq:EL-density} and the definition of $\rho_{\mathrm{eq}}$ as the density of $\mu_{\mathrm{eq}}$ translate into the pair of boundary relations
\begin{align}
    \omega_+(x) + \omega_-(x) &= V'(x)\,,
    \label{eq:resolvent-sum}\\
    \omega_-(x) - \omega_+(x) &= 2\pi i\, \rho_{\mathrm{eq}}(x)\,,
    \label{eq:resolvent-difference}
\end{align}
for $x \in [A,B]$, where $\pm$ denote boundary values from the upper and lower half-planes. Equation \eqref{eq:resolvent-sum} is an additive scalar jump problem, which is solved by dividing through by a function with a compensating multiplicative jump. Set
\begin{equation}
    \mathcal{R}(z) = \sqrt{(z-A)(z-B)}\,,
\end{equation}
with the branch fixed by $\mathcal{R}(z) \sim z$ as $z \to \infty$, so that $\mathcal{R}_+(x) = -\mathcal{R}_-(x)$ on $(A,B)$. Then
\begin{equation}
    \frac{\omega_+(x)}{\mathcal{R}_+(x)}
    - \frac{\omega_-(x)}{\mathcal{R}_-(x)}
    = \frac{\omega_+(x) + \omega_-(x)}{\mathcal{R}_+(x)}
    = \frac{V'(x)}{\mathcal{R}_+(x)}\,,
\end{equation}
so $\omega/\mathcal{R}$ has an additive jump given entirely by known data. Since $\omega(z)/\mathcal{R}(z) = O(z^{-2})$ at infinity, the Cauchy integral formula gives
\begin{equation}
    \frac{\omega(z)}{\mathcal{R}(z)}
    = \frac{1}{2\pi i}\int_A^B
    \frac{ds}{s-z}\,\frac{V'(s)}{\mathcal{R}_+(s)}\,,
    \label{eq:resolvent-solution}
\end{equation}
with no additive entire piece. Deforming the contour to encircle $[A,B]$ and reading off the density from \eqref{eq:resolvent-difference} yields \eqref{eq:vprimee}, namely
\begin{equation}
    \rho_{\mathrm{eq}}(x) = \frac{1}{2\pi}h(x)\sqrt{(x-A)(B-x)}\,,
    \qquad
    h(x) = \frac{1}{2\pi i}\oint_{\Gamma}
    \frac{V'(s)\, ds}{(s-x)\sqrt{(s-A)(s-B)}}\,.
\end{equation}

\noindent
The band edges are fixed by the behaviour of \eqref{eq:resolvent-solution} at infinity. Since $\mu_{\mathrm{eq}}$ is a probability measure,
\begin{equation}
    \omega(z) = \frac{1}{z} + O(z^{-2})\,,
    \qquad z \to \infty\,,
\end{equation}
with no $O(z^{0})$ term. Expanding \eqref{eq:resolvent-solution} and demanding the vanishing of the $z^{0}$ coefficient and the normalisation of the $z^{-1}$ coefficient produces the two conditions \eqref{eq:endpoint-conditions}. For $V(x) = x^2/2$ these are solved by $A = -2$, $B = 2$, giving $h(x) = 1$ and the Wigner semicircle
$\rho_{\mathrm{eq}}(x) = \frac{1}{2\pi}\sqrt{4-x^2}$.

\subsection{The Airy parametrix at a soft edge}
\label{app:airy}

Near a generic endpoint the equilibrium density vanishes as a square root, \eqref{eq:soft-edge}, and consequently the phase $\phi$ of \eqref{eq:phi-def} has a $3/2$-power zero. This sets the local scale. In the disc $D_B$ we introduce the conformal change of variable
\begin{equation}
    \zeta(z) = \left(\frac{3n}{2}\int_B^z
    \sqrt{(s-B)(s-A)}\;\frac{h(s)}{2}\, ds\right)^{2/3}\,,
    \label{eq:airy-variable}
\end{equation}
which is analytic and injective on $D_B$ for $\delta$ small enough and satisfies $\zeta(z) \sim C n^{2/3}(z-B)$ near $B$ with $C > 0$. The exponent $2/3$ arises from inverting the $3/2$-power zero of $\phi$, and the factor $n^{2/3}$ from the $n$ in the varying weight. The width of the region in which the Airy behaviour is resolved is therefore $|z - B| \sim n^{-2/3}$.\\

\noindent
The local parametrix is given by
\begin{equation}
    P^{(B)}(z) = E^{(B)}_n(z)\,
    \Psi_{\mathrm{Ai}}(\zeta(z))\,
    e^{\frac{2}{3}n\zeta^{3/2}\sigma_3}\,,
    \label{eq:airy-parametrix}
\end{equation}
where $\Psi_{\mathrm{Ai}}$ is the standard Airy model solution, assembled sector by sector in the cut $\zeta$-plane from $\mathrm{Ai}(\zeta)$, $\mathrm{Ai}(\varpi\zeta)$, $\mathrm{Ai}(\varpi^2\zeta)$ with $\varpi = e^{2\pi i/3}$, so as to reproduce exactly the jump structure of $S_n$ inside $D_B$ \cite{deift1999orthogonal}. The prefactor $E^{(B)}_n$ is analytic in $D_B$ and is fixed by the matching requirement
\begin{equation}
    P^{(B)}(z)\big(P^{(\infty)}(z)\big)^{-1}
    = \mathbb{I} + O(n^{-1})\,,
    \qquad |z - B| = \delta\,,
    \label{eq:matching}
\end{equation}
which is possible because the large-$\zeta$ asymptotics of the Airy functions reproduce, at leading order, the fourth-root behaviour of the Szeg\H{o} function $\gamma$ in \eqref{eq:szego}. An identical construction with the reflected local variable applies at $A$.

\subsection{The Bessel parametrix at a hard edge}
\label{app:bessel}

For a weight of the form \eqref{eq:hard-edge-weight}, $w(x) = (x-A)^{\alpha}e^{-nV(x)}$ with $\alpha > -1$, the algebraic factor survives the large-$n$ limit and modifies the local behaviour at $A$ in that the equilibrium density diverges as $\rho_{\mathrm{eq}}(x) \sim \tilde c_A (x-A)^{-1/2}$ rather than vanishing. The phase now has a $1/2$-power rather than $3/2$-power zero, and the appropriate local variable is
\begin{equation}
    \tilde\zeta(z) \sim \tilde C\, n^{2}(z-A)\,,
\end{equation}
so the hard edge is resolved on the much narrower scale $|z-A| \sim n^{-2}$. The parametrix is built from the modified Bessel functions $I_{\alpha}, K_{\alpha}$,
\begin{equation}
    P^{(A)}(z) = \tilde E_n(z)\,
    \Psi^{(\alpha)}_{\mathrm{Bes}}(\tilde\zeta(z))\,
    e^{2n\tilde\zeta^{1/2}\sigma_3}\,,
    \label{eq:bessel-parametrix}
\end{equation}
with $\Psi^{(\alpha)}_{\mathrm{Bes}}$ the standard Bessel model solution of order $\alpha$ and $\tilde E_n$ fixed by matching on $\partial D_A$. The matching mismatch is again $O(n^{-1})$, but its residue now depends explicitly on $\alpha$, and this dependence propagates into the $n^{-2}$ coefficient of the recurrence coefficients. For the classical Jacobi weights on $[A,B]$, where both endpoints are hard, the resulting expansion \eqref{eq:jacobi-check} can be checked term by term against the exactly known Jacobi recurrence coefficients.

\subsection{Small-norm analysis and the correction series}
\label{app:small-norm}

The final transformation compares the exact solution to the assembled approximation,
\begin{equation}
    R_n(z) = S_n(z) \times
    \begin{cases}
    \big(P^{(B)}(z)\big)^{-1}, & z \in D_B\,,\\[1mm]
    \big(P^{(A)}(z)\big)^{-1}, & z \in D_A\,,\\[1mm]
    \big(P^{(\infty)}(z)\big)^{-1}, & \text{elsewhere}\,.
    \end{cases}
    \label{eq:R-def}
\end{equation}
By construction $R_n$ has no jump on $(A,B)$ since the constant jump of $S_n$ is cancelled exactly by $P^{(\infty)}$, and no jump inside the discs, where it is cancelled by the local parametrices. What remains are jumps on $\partial D_A \cup \partial D_B$, where the mismatch \eqref{eq:matching} is $\mathbb{I} + O(n^{-1})$, and on the portions of $\Gamma^{\pm}$ and $\mathbb{R}\setminus[A,B]$ outside the discs, where they are $\mathbb{I} + O(e^{-cn})$. The disc boundaries therefore dominate.\\

\noindent
Since all jumps are uniformly close to the identity, the standard small-norm theory for Riemann--Hilbert problems applies. Specifically, $R_n$ exists for $n$ sufficiently large, is unique, and admits an asymptotic expansion
\begin{equation}
    R_n(z) = \mathbb{I} + \frac{R^{(1)}(z)}{n}
    + \frac{R^{(2)}(z)}{n^2} + \cdots\,,
    \label{eq:R-expansion}
\end{equation}
uniformly in $z$, in which each $R^{(k)}$ is computed by iterated Cauchy integrals of the mismatch over $\partial D_A \cup \partial D_B$. Explicitly, writing the jump as $\mathbb{I} + \Delta_n$ with $\Delta_n = O(n^{-1})$,
\begin{equation}
    R^{(1)}(z) = \frac{1}{2\pi i}
    \oint_{\partial D_A \cup \partial D_B}
    \frac{n\Delta_n(s)}{s - z}\, ds\,,
\end{equation}
and higher orders follow by iteration.\\

\noindent
Unwinding the chain \eqref{eq:transformation-chain} expresses $Y_{1,n}$ in terms of the expansion coefficients of $R_n$ and $P^{(\infty)}$, and \eqref{eq:bn}, \eqref{eq:an} then give the recurrence coefficients. The result at a soft edge is \eqref{eq:soft-corrections}. The vanishing of the $O(n^{-1})$ term there deserves comment; the residues of $R^{(1)}$ at the two endpoints enter the combination extracted by \eqref{eq:bn} with opposite signs, so their contributions cancel and the first surviving correction is $O(n^{-2})$. The cancellation is specific to the recurrence coefficients since individual entries of $Y_{1,n}$ do receive $O(n^{-1})$ corrections, and it fails when the two endpoints are not of the same type, for instance when one is soft and the other hard.

\section{Local analysis at the gap-closing transition}
\label{app:critical}

This appendix supplies the computations summarised in Section \ref{sec:critical-krylov}. We derive the matching constants $c_0, c_1$ of \eqref{eq:local-variables}, compare the Airy and Painlev\'e~II expansion orders, and extract the $n^{-1/3}$ correction to the recurrence coefficients. Throughout we work with the symmetric quartic, for which every quantity is explicit. The general case differs only in the value of the single local constant $\kappa$. We follow \cite{bleher2003double, claeys2006universality}, to which we refer for the rigorous statements.

\subsection{The symmetric quartic at criticality}
\label{app:quartic}

Take
\begin{equation}
    V'(x) = g x^3 + t x\,, \qquad g > 0\,,
    \label{eq:quartic-potential}
\end{equation}
with $t$ the deformation parameter. In the one-cut phase the equilibrium measure is supported on $[-b,b]$ with
\begin{equation}
    \rho_{\mathrm{eq}}(x) = \frac{1}{2\pi}h(x)\sqrt{b^2 - x^2}\,,
    \qquad
    h(x) = g x^2 + \Big(t + \tfrac12 g b^2\Big)\,,
    \label{eq:quartic-onecut}
\end{equation}
obtained from \eqref{eq:vprimee}. The normalisation condition $\int\rho_{\mathrm{eq}} = 1$ then reads
\begin{equation}
    \frac{g b^4}{16} + \Big(t + \tfrac12 g b^2\Big)\frac{b^2}{4} = 1\,.
    \label{eq:quartic-normalisation}
\end{equation}
The transition occurs where $h$ develops its double zero at the interior point $x=0$, that is at $h(0) = 0$. Together with \eqref{eq:quartic-normalisation} this gives
\begin{equation}
    t_c = -2\sqrt{g}\,, \qquad b_c = 2 g^{-1/4}\,,
    \label{eq:quartic-critical}
\end{equation}
and at that point $h(x) = g x^2$, so that
\begin{equation}
    \rho_{\mathrm{eq}}(x)\Big|_{t=t_c}
    = \frac{g}{2\pi}\,x^2\sqrt{b_c^2 - x^2}
    \;\sim\; \kappa\, x^2\,,
    \qquad
    \kappa = \frac{g\, b_c}{2\pi} = \frac{g^{3/4}}{\pi}\,,
    \label{eq:quartic-kappa}
\end{equation}
which is \eqref{eq:quadratic-vanishing} with $\omega_c = 0$. For $g=1$ these are $t_c = -2$, $b_c = 2$ and $\kappa = 1/\pi$.\\

\noindent
On the two-cut side, $t < t_c$, the support is $[-b,-a]\cup[a,b]$ with
\begin{equation}
    \rho_{\mathrm{eq}}(x) = \frac{g}{2\pi}\,|x|
    \sqrt{(x^2 - a^2)(b^2 - x^2)}\,,
    \qquad
    a^2 + b^2 = -\frac{2t}{g}\,,
    \qquad
    \frac{g\,(b^2-a^2)^2}{16} = 1\,.
    \label{eq:quartic-twocut}
\end{equation}
Solving the two conditions gives $a^2 = -\big(t + 2\sqrt{g}\big)/g$, so the inner edges open as
\begin{equation}
    a = \sqrt{\frac{t_c - t}{g}}\,,
    \qquad
    \epsilon_{\mathrm{gap}} = 2a\,.
    \label{eq:quartic-gap}
\end{equation}
Setting $\sigma = (t-t_c)/g$ we have $\sigma = -a^2$, and hence
\begin{equation}
    \sigma = -\tfrac14\epsilon_{\mathrm{gap}}^{\,2}\,,
    \label{eq:app-sigma-gap}
\end{equation}
which is \eqref{eq:sigma-gap}. This is the sense in which $\sigma$, not the gap width, is the natural deformation variable as it is linear in the coupling $t$, and the gap opens as its square root. Expanding \eqref{eq:quartic-twocut} near the origin for small $a$ gives $\rho_{\mathrm{eq}} \sim \kappa\,(x^2 - a^2) = \kappa(x^2 + \sigma)$, so the two phases are described by the single local model
\begin{equation}
    \rho_{\mathrm{eq}}(u) \;\sim\; \kappa\,\big(u^2 + \sigma\big)\,,
    \qquad u = \omega - \omega_c\,,
    \label{eq:local-model}
\end{equation}
with $\sigma<0$ in the gapped phase and $\sigma>0$ in the merged one.

\subsection{Matching the phase: the constants $c_0$ and $c_1$}
\label{app:matching-constants}

Integrating \eqref{eq:local-model} through \eqref{eq:phi-def} gives the phase
\begin{equation}
    \phi(u) - \phi(0) = -2\pi\int_0^u \rho_{\mathrm{eq}}
    = -2\pi\kappa\left(\frac{u^3}{3} + \sigma u\right),
    \label{eq:app-phase}
\end{equation}
and the exponent controlling the jumps of $S_n$ near $\omega_c$ is therefore $n\phi$. The Painlev\'e~II $\Psi$-function of \eqref{eq:PII-lax} carries the phase
\begin{equation}
    \theta(\zeta,s) = \tfrac43\zeta^3 + s\,\zeta\,.
    \label{eq:PII-phase}
\end{equation}
The local variable and the double-scaling parameter are fixed by requiring these to agree, $n|\phi(u)-\phi(0)| = \theta(\zeta,s)$, as an identity in $u$. Writing $\zeta = c_1 n^{1/3} u$ and matching the cubic and linear terms separately,
\begin{align}
    \tfrac43 c_1^3\, n\, u^3 &= \tfrac23\pi\kappa\, n\, u^3
    &&\Longrightarrow\quad c_1 = \Big(\frac{\pi\kappa}{2}\Big)^{1/3},
    \label{eq:c1-match}\\[1mm]
    s\, c_1 n^{1/3} u &= 2\pi\kappa\, n\,\sigma\, u
    &&\Longrightarrow\quad
    s = \frac{2\pi\kappa}{c_1}\, n^{2/3}\sigma
    \;\equiv\; c_0\, n^{2/3}\sigma\,.
    \label{eq:c0-match}
\end{align}
Eliminating $c_1$ from the second line gives
\begin{equation}
    c_0 = \frac{2\pi\kappa}{(\pi\kappa/2)^{1/3}}
    = 2^{4/3}\,(\pi\kappa)^{2/3}\,,
    \label{eq:c0-closed}
\end{equation}
so both constants are determined by $\kappa$ alone. For the symmetric quartic with $g=1$, where $\kappa = 1/\pi$ by \eqref{eq:quartic-kappa}, these evaluate to $c_1 = 2^{-1/3} \approx 0.7937$ and $c_0 = 2^{4/3} \approx 2.5198$.\\

\noindent
Two features of \eqref{eq:c1-match}--\eqref{eq:c0-match} are worth isolating. The exponent $1/3$ in $\zeta \sim n^{1/3}u$ is forced by the cubic phase; a phase $n u^3$ is $O(1)$ when $u \sim n^{-1/3}$, which is the width of the region in which the local model must be used. And the exponent $2/3$ in $s \sim n^{2/3}\sigma$ follows from the linear term, since $\sigma$ must enter multiplied by $n$ and divided by the $n^{1/3}$ already absorbed into $\zeta$. The two exponents are not independent; they are the cubic and linear gradings of the same phase.

\subsection{Expansion orders: Airy versus Painlev\'e~II}
\label{app:expansion-orders}

The difference between the generic $O(n^{-2})$ relaxation and the critical $O(n^{-1/3})$ is entirely a difference in the order of the local expansion, and it is clearest set side by side.\\

\noindent
\textbf{Airy, at a generic soft edge.} The local variable is $\zeta \sim n^{2/3}(z-B)$ by \eqref{eq:airy-variable}. The Airy model solution has the large-$\zeta$ expansion
\begin{equation}
    \Psi_{\mathrm{Ai}}(\zeta)\, e^{\frac23\zeta^{3/2}\sigma_3}
    = \zeta^{-\sigma_3/4}\,\frac{1}{\sqrt2}
    \begin{pmatrix} 1 & i \\ i & 1\end{pmatrix}
    \left(\mathbb I + \frac{A_1}{\zeta^{3/2}}
    + \frac{A_2}{\zeta^{3}} + \cdots\right),
    \label{eq:airy-expansion}
\end{equation}
proceeding in powers of $\zeta^{-3/2}$, with $A_1$ containing the familiar coefficients $\pm 5/48$, $\pm 7/48$. On $\partial D_B$ we have $|\zeta| \sim n^{2/3}\delta$, so the mismatch is
\begin{equation}
    \Delta_n^{\mathrm{Ai}} \;\sim\; |\zeta|^{-3/2}
    \;\sim\; \big(n^{2/3}\big)^{-3/2} = n^{-1}\,,
    \label{eq:airy-mismatch}
\end{equation}
giving $R^{(1)} = O(n^{-1})$. That this produces $O(n^{-2})$ rather than $O(n^{-1})$ in the recurrence coefficients is the endpoint cancellation discussed in Appendix \ref{app:small-norm} with the two endpoints contributing residues of opposite sign to the combination extracted by \eqref{eq:bn}.\\

\noindent
\textbf{Painlev\'e~II, at the critical point.} The local variable is $\zeta \sim n^{1/3}u$ by \eqref{eq:local-variables}. The $\Psi$-function of the Flaschka--Newell Lax pair \eqref{eq:PII-lax} has the large-$\zeta$ expansion
\begin{equation}
    \Psi(\zeta,s)\, e^{i\theta(\zeta,s)\sigma_3}
    = \mathbb I + \frac{\Psi_1(s)}{\zeta} + \frac{\Psi_2(s)}{\zeta^2}
    + O(\zeta^{-3})\,,
    \label{eq:app-PII-expansion}
\end{equation}
proceeding in powers of $\zeta^{-1}$, with
\begin{equation}
    \Psi_1(s) = \frac{i}{2}
    \begin{pmatrix}
    -\displaystyle\int_{\infty}^{s}\! q(r)^2\, dr & q(s) \\[2mm]
    q(s) & \displaystyle\int_{\infty}^{s}\! q(r)^2\, dr
    \end{pmatrix}\,.
    \label{eq:Psi1}
\end{equation}
The off-diagonal entries are the Hastings--McLeod solution itself. On $\partial D_c$ we have $|\zeta| = c_1 n^{1/3}\delta$, so the mismatch is
\begin{equation}
    \Delta_n^{\mathrm{PII}} \;\sim\; |\zeta|^{-1}
    \;\sim\; \big(n^{1/3}\big)^{-1} = n^{-1/3}\,,
    \label{eq:app-PII-mismatch}
\end{equation}
which is \eqref{eq:critical-mismatch}. Both effects push in the same direction. The critical region is wider, $n^{-1/3}$ against $n^{-2/3}$, so $|\zeta|$ on the disc boundary is smaller; and the expansion is of lower order, $\zeta^{-1}$ against $\zeta^{-3/2}$, so a given $|\zeta|$ buys less suppression. The product of the two is the gap between $n^{-1}$ and $n^{-1/3}$.

\subsection{Extraction of the recurrence coefficients}
\label{app:critical-extraction}

It remains to pass from $R^{(1)}$ to $b_n$. By the small-norm theory of Appendix \ref{app:small-norm}, with jumps $\mathbb I + \Delta_n$ on $\partial D_c$,
\begin{equation}
    R_n(z) = \mathbb I + \frac{1}{2\pi i}\oint_{\partial D_c}
    \frac{\Delta_n(\varsigma)}{\varsigma - z}\, d\varsigma
    + O\big(\|\Delta_n\|^2\big)\,.
    \label{eq:app-R-cauchy}
\end{equation}
Since $\Delta_n$ extends meromorphically inside $D_c$ with a simple pole at $\omega_c$, the contour integral is fixed by the residue there and is independent of $\delta$, as expected. Writing
\begin{equation}
    \Delta_n(\varsigma) = \frac{\Lambda_n}{c_1 n^{1/3}(\varsigma - \omega_c)}
    + O(n^{-2/3})\,,
    \qquad
    \Lambda_n = E^{(\mathrm{crit})}_n \Psi_1(s)\,
    \big(E^{(\mathrm{crit})}_n\big)^{-1}\,,
    \label{eq:app-Delta-pole}
\end{equation}
the residue gives
\begin{equation}
    R_n(z) = \mathbb I + \frac{\Lambda_n}{c_1 n^{1/3}\,(z - \omega_c)}
    + O(n^{-2/3})\,,
    \label{eq:app-R-explicit}
\end{equation}
whose large-$z$ expansion has first coefficient
\begin{equation}
    R^{(1)}_{\infty} \;=\; \lim_{z\to\infty} z\big(R_n(z) - \mathbb I\big)
    \;=\; \frac{\Lambda_n}{c_1 n^{1/3}}\,.
    \label{eq:app-R1-infty}
\end{equation}
Unwinding $Y_n \to T_n \to S_n \to R_n$ and applying \eqref{eq:bn} in the form used in Appendix \ref{app:small-norm}, the off-diagonal entries of $R^{(1)}_{\infty}$ enter $b_n$ additively at leading order, so that
\begin{equation}
    \frac{b_n - b_\infty}{b_\infty}
    = \frac{\big(R^{(1)}_{\infty}\big)_{12}
    + \big(R^{(1)}_{\infty}\big)_{21}}{2}
    + O(n^{-2/3})\,.
    \label{eq:app-bn-from-R}
\end{equation}
By \eqref{eq:Psi1} both off-diagonal entries of $\Psi_1$ are proportional to $q(s)$, and the conjugation by $E^{(\mathrm{crit})}_n$ preserves this while contributing the phase that alternates with the parity of $n$. Collecting constants into $d_1$,
\begin{equation}
    b_n = b_\infty\left(1 + \frac{(-1)^n d_1\, q(s)}{n^{1/3}}
    + O(n^{-2/3})\right),
    \label{eq:app-double-scaling-bn}
\end{equation}
which is \eqref{eq:double-scaling-bn}.\\

\noindent
The alternating term here deserves a word, since it is the one feature not fixed by the scaling analysis alone. At the transition $\Omega\to\tfrac12$, and the oscillatory factor $e^{2\pi i n\Omega}$ of \eqref{eq:gap-jump} degenerates to $(-1)^n$. The prefactor $E^{(\mathrm{crit})}_n$ inherits this from the matching to $P^{(\infty)}$, which still carries the two-cut structure outside $D_c$. The $(-1)^n$ in \eqref{eq:app-double-scaling-bn} is therefore the surviving remnant of the theta-function quasiperiodicity of Section \ref{sec:multi-cut}, evaluated at the rational point $\Omega=\tfrac12$, rather than an independent feature of the local analysis. For an asymmetric merge, where $\Omega_c\neq\tfrac12$, the same argument would give a modulation at frequency $\Omega_c$; we have not verified that the Painlev\'e~II local model survives in that case.

\subsection{Consistency: recovering the two-cut staggering}
\label{app:critical-consistency}

As a check of \eqref{eq:app-double-scaling-bn}, consider the limit $s\to-\infty$, deep in the gapped phase. By \eqref{eq:HM-minus-infinity},
\begin{equation}
    q(s) \sim \sqrt{-s/2}
    = \sqrt{\frac{c_0\, n^{2/3}|\sigma|}{2}}
    = n^{1/3}\sqrt{\frac{c_0|\sigma|}{2}}\,,
    \label{eq:app-q-asymptotic}
\end{equation}
so the explicit $n^{-1/3}$ in \eqref{eq:app-double-scaling-bn} is cancelled exactly and
\begin{equation}
    b_n \;\longrightarrow\; b_\infty\left(1 + (-1)^n d_1
    \sqrt{\frac{c_0|\sigma|}{2}}\right).
    \label{eq:app-staggering-limit}
\end{equation}
The correction is an $n$-independent finite even--odd staggering. Using \eqref{eq:app-sigma-gap} this can be expressed as
\begin{equation}
    \frac{\tilde b}{b_\infty}
    = d_1\sqrt{\frac{c_0}{2}}\;\frac{\epsilon_{\mathrm{gap}}}{2}\,,
    \label{eq:app-amplitude-gap}
\end{equation}
so the staggering amplitude is linear in the gap width. This is exactly the structure found independently in Section \ref{subsec:solvable} for the SSH chain, where \eqref{eq:ssh-exact-limits} gives $\tilde b = |t_1 - t_2|/2$, again half the gap. The Painlev\'e~II formula therefore reproduces the two-cut staggering in the appropriate limit, with the correct linear dependence on the gap, and \eqref{eq:app-double-scaling-bn} is the analytic continuation of that staggering through the point at which the gap closes.\\

\noindent
In the opposite $s\to+\infty$ limit the Hastings--McLeod solution decays as $q(s)\to\mathrm{Ai}(s)$, so the staggering switches off exponentially and $b_n\to b_\infty$ monotonically, recovering the single-cut result \eqref{eq:leading-asymptotics} on the merged support. The two limits bracket the transition, and \eqref{eq:app-double-scaling-bn} interpolates between them on the scale $\sigma\sim n^{-2/3}$, or equivalently $\epsilon_{\mathrm{gap}}\sim n^{-1/3}$.
\bibliographystyle{JHEP}
\bibliography{biblio.bib}
\end{document}